\documentstyle[prb,aps,multicol,epsf]{revtex}
\draft
\def\beginwide{\end{multicols} \vspace*{-0.5cm} \noindent\rule{3.5in}{.1mm}
\rule{.1mm}{5mm} \widetext \medskip }
\def\beginwidetop{
        \end{multicols} \vspace*{-0.5cm} \noindent
        \widetext \medskip }
\def\endwide{\hspace*{3.35in}~\rule[-5mm]{.1mm}{5mm}\rule{3.5in}{.1mm}
        \begin{multicols}{2} \vspace*{-1.0cm} \noindent }
\def\endwidebottom{
        \begin{multicols}{2} \vspace*{-1.0cm} \noindent }

\def \begincol{\begin{multicols}{2}} \def\endcol{\end{multicols} }

\newcommand{\epp}{\epsilon_+}
\newcommand{\ep}{\epsilon}
\newcommand{\mc}{m_{\parallel}}
\newcommand{\mab}{m_{\perp}}
\newcommand{\epm}{\epsilon_-}
\newcommand{\ah}{|\alpha_H|}

\newcommand{\bp}{{\bf p}}
\newcommand{\br}{{\bf r}}
\newcommand{\tT}{{\tilde{T}}}
\newcommand{\bk}{{\bf k}}
\newcommand{\bq}{{\bf q}}
\newcommand{\tcsix}{\mbox{$\tilde{c}_{66}$}}
\newcommand{\trhos}{\mbox{${\tilde{\rho}_s}$}}
\newcommand{\xic}{\xi_{\parallel}}
\newcommand{\xiab}{\xi_{\perp}}

\newcommand{\tk}{\tilde{k}}
\newcommand{\ar}{\tilde{\alpha}}
\newcommand{\ellc}{\ell_{\parallel}}
\newcommand{\ellab}{\ell_{\perp}}
\newcommand{\betak}{\beta_\kappa}
\newcommand{\et}{{\it et~al.} }
\newcommand{\dif}{\partial}

\begin{document}

\title{
Finite size effects as the explanation of ``freezing'' in vortex liquids}
\author{ S-K. Chin and M. A. Moore}
\address{
Theory Group,
Department of Physics and Astronomy,\\
University of Manchester, M13 9PL, UK.\\}
\date{\today}

\maketitle
\begin{abstract}
We investigate the effect of thermal fluctuations on the
(mean-field) Abrikosov phase.
The lower critical dimension of the superconducting phase
is three,
indicating the absence of the Abrikosov phase for dimensions $d\le 3$.
Within the $d=3$ vortex liquid,
the phase correlation length
$\ellc$ along the magnetic field direction grows exponentially
rapidly as the temperature is lowered.
For a finite bulk system,
there is a 3D-2D crossover effect when $\ellc$
becomes comparable to the sample thickness.
Such a crossover effect
takes place over a very narrow temperature interval and
mimics the ``first order transition'' seen in experiments on clean
${\rm YBa_2Cu_3O_{7-\delta}}$ (YBCO) and ${\rm Bi_2Sr_2CaCu_2O_8}$ (BSCCO)
crystals.
We calculate the jumps in the entropy,
magnetization and specific heat
due to the crossover and find reasonably
good agreement with experiments on both YBCO and BSCCO.
\end{abstract}


\begincol
\section{Introduction}
Abrikosov's mean-field
treatment of conventional Type II superconductors in a magnetic field
is very accurate \cite{Abrikosov57} because of their extremely
narrow critical regime.
In this approximation, the superconducting triangular
vortex crystal melts into
a resistive vortex liquid
via a continuous phase transition.
However because of the strong thermal fluctuations
about the mean-field solution in high $T_c$ superconductors (HTSC) such as
YBCO and BSCCO,
the nature of the transformation between
the superconducting and the normal phase (the vortex liquid),
and even the very existence
of the mixed phase itself
has became an issue of great interest and complexity.

Evidence for a first order melting phase transition from vortex
crystal to vortex liquid has been found in recent magnetization
measurements on clean crystals of YBCO \cite{Liang96,Welp96} and
BSCCO\cite{Zeldov95}.
The magnetization jump $\Delta M$ associated with
the melting is found to be
about $3\times 10^{-5}$T at $4$T for YBCO and
$4\times 10^{-5}$T at $5\times 10^{-3}$T for BSCCO.
By measuring the position of the phase boundary in the $H$-$T$ plane,
and assuming
a first order phase transition, one can obtain
the entropy jump per vortex per CuO layer $\Delta S$ via the
Clausius-Clapeyron equation:
\begin{eqnarray}
\frac{\Delta S}{\Delta M}=-\frac{s\Phi_0}{B} \frac{ d H_m(T)}{dT}~,
\label{eqn_claperon}
\end{eqnarray}
where $B$, $s$ and $H_m(T)$ are
magnetic induction along the $c$-axis,
layer spacing and melting line respectively and
$\Phi_0$ is the flux quantum.
For YBCO, $\Delta S$ is calculated to be $0.6k_B$/layer/vortex
at $4$T\cite{Welp96}.
For BSCCO, $\Delta S$ is calculated to be much higher:
about $2k_B$/layer/vortex at $1\times 10^{-4}$T, and rapidly growing
as the temperature approaches $T_c$\cite{Zeldov95}.
Recently,
both the magnetization jump {\em and} the latent heat
have been measured on the same YBCO crystal.
While Schilling \et \cite{Schilling96} confirmed that the
jumps in entropy/vortex/layer and magnetization satisfied
the Clausius-Clapeyron equation, Junod \et\cite{Junod97} found less
satisfactory agreement,
possibly due to sample inhomogeneities.
Both sets of authors report a disappearance of the entropy jump in small
fields, which again probably indicates that the experimental data is
being affected by sample artifacts.

Recently reported numerical simulations also favor
a first order melting transition.
Monte Carlo simulations using variants of the 3D XY model
give a jump in the entropy\cite{Hetzel92,Chen97,Nguyen97}.
Sasik and Stroud's 3D Monte Carlo simulation within the lowest Landau level
(LLL) approximation\cite{Sasik95b} (when corrected for an erroneous
definition of
$\Delta S$) and that of Hu and MacDonald\cite{Hu97} yield estimates of
$\Delta S\approx 0.6 k_B$/layer/vortex,
in good agreement with YBCO experiments.
However, the results from these simulations
in the LLL scheme should be treated with caution:
they were performed using quasi-periodic boundary
conditions, which imposes an effective (spurious) pinning potential
on the vortex motion\cite{Dodgson97}.
A good example of the problem of using the quasi-periodic
boundary condition is the case of
two-dimensional (2D) (thin film) simulations within the
LLL approximation:
authors who use quasi-periodic boundary
conditions see first order vortex melting
\cite{Sasik93,Sasik94,Tesanovic91,Kato93,Hu93}, whereas
simulations in which the 2D vortices move on the surface of
a sphere \cite{Oneill93,Dodgson97}, which
involves no spurious pinning potential,
see no phase transition at all!
There is also no experimental evidence that thin film superconductors
undergo a first order melting phase transition.
A detailed discussion of this topic has been given in
Ref.\,\onlinecite{Dodgson97}.

On the theoretical front, it has been thought for a long time
that
the Abrikosov lattice will melt into a vortex liquid, and that the phase
transition will be first order\cite{Brezin85}.
However, there is as yet no detailed melting theory. Over the years,
many theoretical investigations
based on the melting scenario have relied upon the Lindemann criterion:
when the spatial fluctuations of a vortex
due to thermal excitations become some fraction $c_L$ of the
vortex lattice spacing, the lattice melts. The Lindemann number
$c_L$ is usually in the range 0.2--0.4\cite{Blatter94}.
It has also been suggested that the mechanism behind the apparent
first order
phase transition is the decoupling of the vortex-lines to
pancakes\cite{Daemen93}. However, this does not explain the
disappearance at the transition of the crystalline order
seen in
neutron scattering experiments \cite{Cubitt93}.

On the other hand, it is well known that the mean-field Abrikosov
solution is unstable against
long wavelength thermal
excitation of the shear modes of the vortex lattice
in both the physical
dimensions
$d=2$ and $3$ \cite{Maki71,Moore89}.
By calculating the off-diagonal long range order (ODLRO)
in the low temperature regime,
it has been shown that the lower critical dimension $d_{LC}$
of the mixed phase is three for extreme Type II superconductors ($\kappa>>1$)
in an external
magnetic field\cite{Moore92}.
(The lower critical dimension of an ordered phase is the spatial dimension at
and below which a system can no longer sustain the long range order
associated with that phase at non-zero temperature).
It was suggested therefore that thermal fluctuations will
modify the mean-field
phase diagram as follows: for $d=2$, the normal vortex liquid is
the only thermodynamic phase.
For $d=3$, there are only the Meissner and the normal
vortex liquid phases.
This theoretical scenario for $d=3$ is seemingly at odds with
the overwhelming experimental and numerical evidence for a first
order melting transition as outlined above,
and hence has attracted little attention.

Recently, however, one of us \cite{Moore97} has proposed that the apparent
first order melting transition of a vortex crystal to a liquid
may be a signature of a finite size effect
rather than a genuine thermodynamic phase transition.
This idea is based on an extension of Refs.
\onlinecite{Moore89} and \onlinecite{Moore92}.
In this picture for $d=3$, there exits only one true thermodynamic phase,
the vortex liquid phase,
characterized by two length scales
$\ellc$ and $\ellab$;
$\ellc$ measures the
phase correlation length along the magnetic field direction,
and $\ellab$ is the range of the
(short-range) crystalline order.
Both $\ellc$ and $\ellab$ are growing exponentially rapidly
as the temperature is lowered,
but they
only become infinite at zero temperature.
There is no phase transition
to an ordered phase at any finite temperature.
This scenario is that of zero temperature
scaling\cite{Oneill93,Moore92,Moore97}.
For a finite system, the rapid growth of $\ellc$ and $\ellab$
as the temperature is lowered has profound consequences.
For a bulk sample with a slab geometry and the field along the $c$-axis,
the vortex liquid phase
becomes phase correlated along the field direction upon cooling
when $\ellc$ reaches the sample thickness $L_z$. Then one has
phase correlation right across the sample.
The behavior of the system then crosses sharply over from
that of a 3D vortex liquid to that of a 2D
vortex liquid \cite{decouple}.
Such a crossover effect can explain the sudden drop in the $c$-axis resistivity
\cite{Moore97}.
We will show later that this crossover can
quantitatively explain the apparent jumps in the entropy and
magnetization of the system observed in experiments.
In this simple scenario
there is no melting phase transition
and a vortex crystal phase does not occur.

The rapid growth of the $c$-axis phase correlation
may have already been seen in flux transformer experiments
on clean YBCO crystals\cite{Lopez96}.
Right at the point where the apparent first order melting
transition occurs,
the voltage difference between
various points on the top and bottom of the sample are as if
the flux lines moved as rigid rods,
indicating phase coherence across the sample thickness.
(We acknowledge that a growing $c$-axis conductivity can also produce the
same behavior without any substantial degree of phase coherence being present).
Recent numerical simulations in frustrated Villain\cite{Nguyen97},
XY \cite{Chen97} and LLL \cite{Kienappel97b} models have
reported
rapidly growing phase coherence along the $c$-axis
upon cooling.
With a conventional first order
phase transition picture,
it is difficult
to explain the presence of such a
growing length scale at the transition.
In addition, the jumps in the magnetization and entropy in
Refs.\,\onlinecite{Welp96} and \onlinecite{Schilling97}
are actually rounded as functions of magnetic field and temperature.
The width of the transition has been calculated
within the crossover approach\cite{Moore97},
and is in good agreement with data. If there were a genuine first order
transition this rounding has to be explained on the basis of a
sample artifact etc.

In this paper, we shall follow the zero temperature scaling idea
and focus on the growing length scale and the role of
finite size effects.
The outline of this paper is as follows:
we start by describing the LLL approximation based on the
the work of Eilenberger \cite{Eilenberger67} and use it to study
anharmonic fluctuations
in the Abrikosov phase.
We then reproduce in Section\,\ref{sec_eqn_of_state} the result that
$d_{LC}$ of the mixed phase is three
\cite{Maki71,Moore89,Moore92,Moore97}
within the framework
of the loop expansion
and estimate
the growth rate of $\ellc$ and $\ellab$ as a function
of temperature in three dimensions.
We find the 3D-2D crossover line by setting $\ellc \approx L_z$,
and compare it to the experimental melting line
with the Ginzburg number as the fitting parameter.
Both YBCO and BSCCO are examined.
In Section \ref{sec_thermodynamics},
we argue that the three dimensional form of the free energy crosses over sharply to
the two dimensional thin-film form,
mimicking the sharp changes in the thermodynamic potential associated
with a first order phase transition. We will calculate
the jumps in the first derivative of the free energy
to find the entropy jump/layer/vortex,
$\Delta S_{cr}$ and the magnetization $\Delta M_{cr}$
and compare them with experiment
(where the jumps are usually
interpretated as being due to a first order phase transition).
We will also demonstrate that $\Delta S_{cr}$ and $\Delta M_{cr}$
satisfy
the Clausius-Clapeyron equation.
With sets of parameters appropriate for YBCO and BSCCO,
we will show that our results are in reasonably
good agreement with experiments.
The effect of weak random disorder, which is always present even in clean
crystals, is investigated in
Section\,\ref{sec_disorder}.
Finally, we will conclude with a summary and discussion in Section
\ref{sec_conclusion}. Most of the details of the (necessarily) complicated
calculations are to be found in Appendices \ref{appendix_shear}--\ref{appendix_i}.

\section{The Model}
\label{sec_model}
We start from the Ginzburg-Landau model with a complex superconducting
order parameter $\Psi$ for a system with spatial dimension $d$ and in
an external magnetic field ${\bf H}_0$ along the $(d-2)$ longitudinal
directions. It is assumed that the system has an effective anisotropy
${\bf m}=(\mab,\mab,{\bf m}_\parallel)$. The free energy functional is
\begin{eqnarray}
{\cal F}=\int d^d r \bigg[ &&\alpha |\Psi|^2
+\sum_{i=1}^d\frac{|(-i\hbar\nabla-2eA_i)\Psi|^2}{2m_i}
\nonumber\\
&&+\frac{1}{2} \beta |\Psi|^4
+\frac{|{\bf B}-\mu_0{\bf H}_0|^2}{2\mu_0}
\bigg],
\label{eqn_gl}
\end{eqnarray}
where $\alpha\propto (T-T_c)$ and $\beta$ is taken to be a constant.
The magnetic induction ${\bf B}=\nabla \times {\bf A}$
is assumed to be uniform inside the
bulk of the system and parallel to ${\bf H}_0$.
This approximation is valid for an extreme Type II
(GL parameter $\kappa>>1 $) superconductor
where the fluctuations in the vector potential ${\bf A}$
are negligible compared to those of the order parameter.
Choosing a Landau gauge ${\bf A}=B(-y,0,0)$
and restricting the fluctuations of the order parameter to the LLL
subspace, the free energy functional is reduced to:
\begin{eqnarray}
{\cal F}_{LLL}=\int d^d r \left[ \alpha_H |\Psi|^2
+\frac{1}{2} \beta_\kappa |\Psi|^4
+\frac{\hbar^2 }{2\mc}
\bigg|\frac{\partial \Psi}{\partial {\bf z}}\bigg|^2
\right],
\label{eqn_gl_lll}
\end{eqnarray}
where $\beta_\kappa=\beta(1-1/2\kappa^2)$ and
$\alpha_H=\alpha+e\hbar \mu_0 H_0/\mc$. $\alpha_H=0$ defines the
mean-field $H_{c2}$ line below which the Abrikosov mean-field solution
can be written as\cite{Eilenberger67}
\begin{eqnarray}
\Psi_0 &=&\alpha_0 \varphi({\bf r}|0),\\
\varphi({\bf r}|0)& =& \left(2\eta\right)^{1/4}
\exp\left(-\frac{y^2}{2\ell^2}\right)
\vartheta_3\left(\frac{\pi(x+iy)}{\ell_0}\bigg|\zeta+i\eta\right).
\end{eqnarray}
$\Psi_0$ has zeros which form a triangular lattice
with the fundamental unit cell spanned by the vectors
$\br_{\rm I}=(1,0)\ell_0$ and $\br_{\rm II}=(\zeta=1/2,\eta=\sqrt{3}/2)\ell_0$.
$\vartheta_3$ is
a Jacobian theta function and $\ell=\sqrt{\hbar/2eB}$ is the magnetic length.
The spacing of the vortices $\ell_0$ is given by the flux quantization condition
$\eta \ell_0^2=2\pi\ell^2$. For a system with volume $V=L_xL_yL_z^{(d-2)}$,
the number of vortices is $N_\phi=L_xL_y/(2\pi\ell^2)$.
$\Psi_0$ minimizes the free energy ${\cal F}_{LLL}$ at the value
${F}_{MF}=-\ah^2 V/2\beta_\kappa\beta_A$,
with $\alpha_0=\sqrt{\ah/\beta_\kappa\beta_A}$, and
$\beta_A=1.1596\dots$ is the
Abrikosov number for the triangular lattice.
Following Eilenberger \cite{Eilenberger67,Maki71},
we construct an orthonormal basis set for the $N_\phi$-fold degenerate
ground states of the operator $(-i\hbar{\bf \nabla}-2e{\bf A} )^2$,
\{$\Psi_{\bp}=e^{i\bq.{\bf z}}\varphi(\br|\br_k)$\}, of
which $\Psi_0$ is a member.
Each basis is labelled by a vector ${\bf p}=(\bk,\bq)$ where $\bq$
is the $(d-2)$ dimensional longitudinal vector
and $\bk$ is a two
dimensional vector confined to the first Brillouin zone (BZ) associated
with the ideal triangular lattice.
The basis states can be generated using the relation: $\varphi({\bf r}|{\bf r}_k)
=e^{ik_x x}\varphi({\bf r}+{\bf r}_k|0)$
for ${\bf r}_k=(x_k+i y_k)\equiv
\ell^2 (k_y-ik_x)$. The normalization of $\varphi({\bf r}|{\bf r}_k)$
is taken to be $\overline{|\varphi({\bf r}|{\bf r}_k)|^2}=1$, where
the overline denotes a spatial average over the unit cell.
In each of the $(d-2)$ longitudinal dimensions,
the allowed values of $q$ are integral multiples of $2\pi/L_z$ and
there are $N_z=L_z/s$ of them, where $N_z$ is the number of layers.
The number of allowed $\br_k$ is just $N_\phi$.
Therefore, the total number of degrees of freedom in a set of \{$\Psi_\bp$\}
is $N_\phi N_z^{(d-2)}$. To simplify notation, we will drop the bold type
on the vectors $\bp$, $\bk$ and $\bq$ henceforth.

Next we set up the standard formalism for perturbation expansion around the
condensed mode\cite{Ma_book}.
Since the functional in Eq.\,(\ref{eqn_gl})
is not translationally invariant,
the thermal average of $\Psi$ is
spatially inhomogeneous and it is convenient
to define a spatially averaged
quantity $\ar$ by $\ar^2=\overline{|<\Psi>|^2}$.
At mean-field level, $\ar=\alpha_0$.
Writing the fluctuating order parameter around
$\Psi_0$ as

\beginwide
\begin{eqnarray}
\Psi
&=&\ar \varphi(\br|0)+\delta\Psi
\\
\delta\Psi&=& \frac{1}{\sqrt{V}}\sum_p 
\left(\frac{I^*(0,0|k,-k)}{|I(0,0|k,-k)|}\right)^{1/2}
c_{p}\exp(iq.z)\varphi(\br|\br_k),
\label{eqn_fluc1}
\\
I(k_1,k_2|k_3,k_4) &=&
\frac{1}{\beta_A A}\int_{\rm cell} d^2 r\varphi^*(\br|\br_{k_1})
\varphi^*(\br|\br_{k_2})\varphi(\br|\br_{k_3})
\varphi(\br|\br_{k_4}).
\label{eqn_i}
\end{eqnarray}
\endwide
Implicit in the definition Eq.\,(\ref{eqn_i}) is the
conservation of momentum $\delta_{k_1+k_2,k_3+k_4}$, and
the transverse integration
is over the primitive unit cell of area $A$.
$I(k_1,k_2|k_3,k_4)$ can be conveniently
expressed in terms
of lattice sums (see Appendix \ref{appendix_i}).
Substituting Eq.\,(\ref{eqn_fluc1}) into Eq.\,(\ref{eqn_gl_lll}), we find that the
free energy to quadratic order in $c_p$
is given by

\begin{eqnarray}
{\cal F}_2 &=&{\cal F}_{MF}+
\frac{1}{2}\sum_p\left( \begin{array}{cc} c_p & c_{-p}^*\end{array}\right)
\left(
\begin{array}{cc}
\Omega_p & \Lambda_p \\  \Lambda_p & \Omega_p \\
\end{array}\right)
\left(
\begin{array}{c} c_p^* \\ c_{-p}\\ \end{array}\right)
\label{eqn_matrix}
\end{eqnarray}
where
\begin{eqnarray}
\Omega_p & =& \left(-\ah+2\betak\ar^2I(0,k|0,k)+\frac{\hbar^2q^2}{2\mc}\right),
\\
\Lambda_p &=& \betak\ar^2|I(0,0|k,-k)|.
\end{eqnarray}
Eq.\,(\ref{eqn_matrix}) can be diagonalized as
\begin{eqnarray}
{\cal F}_2={\cal F}_{MF}+\frac{1}{2}\sum_{p}
\left(E_+|b_p|^2+E_-|a_p|^2\right)
\label{eqn_eigen}
\end{eqnarray}
with the eigenvalues $E_\pm$ and their corresponding normalised
eigenvectors
${\bf e}_\pm$:
\begin{eqnarray}
E_\pm &= &\Omega_p\pm \Lambda_p
=\ah\left[\frac{\ar^2}{\alpha_0^2}
(\ep_\pm+1)-1+\xic^2q^2\right],
\\
{\bf e}_+& =&\frac{1}{\sqrt{2}}\left(\begin{array}{c} 1 \\ 1 \\ \end{array}\right),
~~~~~~~
{\bf e}_-=\frac{1}{\sqrt{2}}\left(\begin{array}{c} 1 \\ -1 \\ \end{array}\right),
\label{eqn_eigenvector}
\end{eqnarray}
where
$\ep_\pm=2I(0,k|0,k)\pm|I(0,0|k,-k)|-1$,
and $\xic=(\hbar^2/2\mc\ah)^{1/2}$
is the mean-field correlation length in each of the $(d-2)$ dimensions.
The form of Eq.\,(\ref{eqn_eigenvector}) implies that
the variable $c_p$ can be written as $c_p=(ia_p+b_p)/\sqrt{2}$
provided $a_{-p}=a_p^*$ and $b_{-p}=b_p^*$ where the
complex variables $a_p$ and $b_p$ measure the amount of soft or hard mode
generated by the thermal fluctuations. The
number of degree of freedom for each mode is
$N_\phi N_z^{(d-2)}$.
The soft mode and hard mode propagators are
obtained to lowest order from Eq.\,(\ref{eqn_eigen})
and are given respectively by
\begin{eqnarray}
G_{a} (p)&=&<a_p a_p^*>=\frac{k_BT}{\ah(\epm+\xic^2 q^2)},
\\
G_{b}(p)&=&<b_p b_p^*>=\frac{k_BT}{\ah(\epp+\xic^2 q^2)}.
\end{eqnarray}
The soft and hard modes are so-called because
of the asymptotic behavior for small $k$: $\epm(k)\sim k^4$
and $\epp(k)\sim 2$
(see Appendix \ref{appendix_i}). The consequences
of this $k$-dependence of $\epm(k)$
for the lower critical dimension $d_{LC}$ of the system
will be discussed in more detail later.

We shall introduce a fictitious source field $J$ into
the LLL functional
in Eq.\,(\ref{eqn_gl_lll}) such that
\begin{eqnarray}
{\cal F'}={\cal F}_{LLL}-\int d^d r
\left( J\varphi(\br|0)\Psi^*+c.c\right).
\label{eqn_gl_lll_j}
\end{eqnarray}
Below $T_c$, $J$ singles out $\Psi_0\propto\varphi(\br|0)$
as the condensed mode.
All results will finally be evaluated at $J=0$.
We rewrite $\delta \Psi$ in terms of
$a_p$ and $b_p$ as:
\begin{eqnarray}
\delta\Psi &=&\frac{b_0}{\sqrt{2V}} \varphi(\br|0)+
\frac{1}{\sqrt{V}}\sum_{p\neq 0} (ia_p+b_p)Q_k e^{iq.z} \varphi(\br|\br_k),
\label{eqn_deltaPsi}
\end{eqnarray}
where
\begin{eqnarray}
Q_k&=& \left(\frac{I^*(0,0|k,-k)}{2|I(0,0|k,-k)|}\right)^{1/2}.
\end{eqnarray}
The term ($b_0$ is real) which renormalizes the amplitude
of the condensed mode $\varphi(\br|0)$ has been separated out
for reasons which will be clear when calculating the equation of state in
Section\,\ref{sec_eqn_of_state}.
The sum $p\neq 0$ implies that
except ($q, k$)=($0,0$), $q$ takes on all momenta in the $(d-2)$
longitudinal space,
and $k$ takes on all the permitted values in the {\em whole}
of the first BZ.
Expanding Eq.\,(\ref{eqn_gl_lll_j}) using Eq.\,(\ref{eqn_deltaPsi}),
we obtain the free energy functional
\beginwide
\begin{eqnarray}
{\cal F'}& =& 
V\betak\beta_A(-\alpha_0^2\ar^2+\ar^4/2)
-2JV(\ar+b_0/\sqrt{2V})
\nonumber \\
& &+\betak\beta_A\left[\sqrt{2V}b_0(-\ar\alpha_0^2+\ar^3)
+\frac{1}{2}b_0^2(-\alpha_0^2+3\ar^2)
+\frac{\ar b_0^3}{\sqrt{2V}}
+\frac{b_0^4}{8V}
\right]
\nonumber \\
& &+
\frac{1}{2}\sum_{p\neq 0}
\ah\left\{
|a_p|^2
\left[\frac{\ar^2}{\alpha_0^2}(\epm(k)+1)-1+\xic^2 q^2\right]
+|b_p|^2\left[\frac{\ar^2}{\alpha_0^2}(\epp(k)+1)-1+\xic^2 q^2\right]
\right\}
\nonumber \\
& &+\frac{\betak\beta_A}{2V}
\left(\frac{1}{2}b_0^2+\sqrt{2V}b_0\ar\right)
\sum_{p\neq 0}\left[\left(\epm(k)+1\right)|a_p|^2+
\left(\epp(k)+1\right)|b_p|^2\right]
\nonumber \\
& &+\frac{2}{V}\betak\beta_A(\ar\sqrt{2V}+b_0)
\sum_{p_i\neq 0}\Big[
F(0,p_1|p_2,p_3)(b_{p_1}^* b_{p_2} b_{p_3}+2a_{p_1}^* b_{p_2} a_{p_3}
-b_{p_1}^*a_{p_2}a_{p_3})\Big],
\nonumber \\
& &+ \frac{\betak\beta_A}{2V}\sum_{p_i\neq 0}
\Big[F(p_1,p_2|p_3,p_4)\left(
a_{p_1}^*a_{p_2}^* a_{p_3} a_{p_4} +b_{p_1}^* b_{p_2}^* b_{p_3} b_{p_4}
+4a_{p_1}^*b_{p_2}^*a_{p_3}b_{p_4}\right)
\nonumber \\
& &~~~~~~~~~~~~~~~~~-2F(p_1,-p_2|p_3,-p_4)b_{p_1}^*b_{p_2}a_{p_3}a_{p_4}^*\Big]
\label{eqn_quartic}
\end{eqnarray}
where
\begin{eqnarray}
F(p_1,p_2|p_3,p_4)=\delta_{q_1+q_2,q_3+q_4}Re[Q_{k_1}^*Q_{k2}^*
Q_{k_3}Q_{k_4}I(k_1,k_2|k_3,k_4)].
\end{eqnarray}
\endwide

In Eq.\,(\ref{eqn_quartic}), the sum of $p_i$ is over the whole cell
subject to the constraints on $a_p$ and $b_p$.
For the practical purpose of perturbation calculation that follows,
it is most convenient to
impose the constraints explicitly and sum $p_i$ over half over the
$d$-dimensional BZ (i.e. $k$ is restricted to half of the
two-dimensional BZ and $q$ takes on all allowed momenta).

The low-temperature free energy functional
Eq.\,(\ref{eqn_quartic}) can be characterised by an
effective temperature
$\tT=\beta_\kappa k_B T/(2\ell^2 \xic^{d-2}\ah^2)$,
which is related to
another popular variable $\alpha_T$
via $\tT=4\pi/|\alpha_{T}|^{3/2}$ for $d=3$.
For $d=2$, we define $\tT_{2D}=\betak k_BT/2\ell^2L_z\ah^2=\pi/|\alpha_{2T}|^2$.
The low and high temperature limits are represented by
$\tT\rightarrow 0$ or $\alpha_T(\alpha_{2T})\rightarrow-\infty$, and
by $\tT\rightarrow\infty$ or
$\alpha_T(\alpha_{2T}\rightarrow \infty)$ respectively.
Also useful in the following discussion is the
definition of $\alpha_T=-(2/Gi)^{1/3} (ht)^{-2/3}(1-h-t)$, where
$t=T/T_{c}$ and $h=B/B_{c2}(0)$.
$Gi$ is the Ginzburg number defined
in Ref.\,\onlinecite{Blatter94}. $T_{c}$ and $B_{c2}(0)$ are the
zero field transition temperature and the linear extrapolation
of $H_{c2}(T)$ to zero temperature respectively.

\section{The Equation of State}
\label{sec_eqn_of_state}

\begin{figure}
\narrowtext
\centerline{
\epsfxsize= 11cm 
\epsfbox{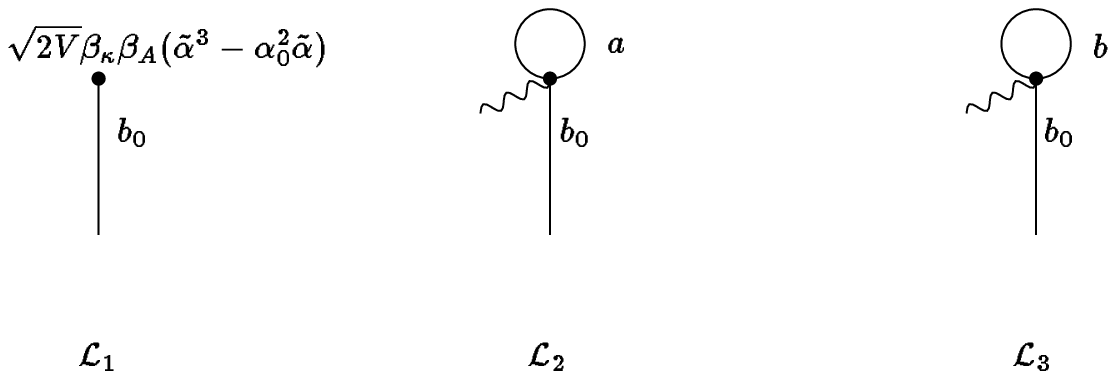}}
\caption{Diagrams for determining the equation of state
via setting \protect $<b_0>={\cal L}_1+{\cal L}_2+{\cal L}_3=0$ up to
${\cal O}(\betak^{1/2})$. The mean-field value of $\ar$ is given by
${\cal L}_1=0$.} 
\label{fig_eqn_of_state} 
\end{figure}

One of the chief aims of this paper is to carry out a
loop expansion
around the mean-field Abrikosov solution---beyond
the Gaussian approximation previously
studied\cite{Eilenberger67,Maki71}---in fact to two loop order.
This loop expansion is well known in the O(n)
model\cite{Wallace_book,Ma_book}.
We stress that it is a systematic perturbative approach
involving no {\it ad hoc} Ansatz (such as was
employed in Ref.\onlinecite{Ruggeri79}).
We shall first calculate the
equation of state by finding what value of $\ar$ makes $<b_0>=0$.
Within the Gaussian approximation, this just corresponds to
putting to zero the
coefficient of $b_0$ in the functional Eq.\,(\ref{eqn_quartic})
(see diagram ${\cal L}_1$ of Fig.\,\ref{fig_eqn_of_state}). This
reproduces the mean-field results $\ar=\alpha_0$.

To one loop order, the equation of state
involves the tadpole diagrams shown in Fig.\,\ref{fig_eqn_of_state}.
Explicit expressions for each diagram are given in Appendix \ref{appendix_ga}.
The soft and hard mode propagators $G_a(p)$ and $G_b(p)$ are labelled
by $a$ and $b$ respectively.
The wavy line in the Feynman diagrams denotes
the order parameter $\Psi_0$.
The thermal averaged value of the amplitude of the
order parameter $\tilde{\alpha}=(\overline{|<\Psi>|^2})^{1/2}$
will be calculated, where
the thermal and spatial averages are
denoted by angular bracket and an overline respectively.
To one loop order the value of $\tilde{\alpha}$ is

\beginwide
\begin{eqnarray}
\tilde{\alpha}^2&=&\alpha_0^2\left\{
1-\frac{\beta_\kappa\beta_A k_B T}{\gamma \ah(2\pi)^d}
\int_{BZ} d^2 k' \int d^{d-2}q' \left[\frac{\epp(k')+1}{\epp(k')/\xic^2+q'^2}+
\frac{\epm(k')+1}{\epm(k')/\xic^2+q'^2}\right]\right\}.
\label{eqn_of_state}
\end{eqnarray}
\endwide
where $\gamma=\hbar^2/\mc$.
For $3<d<4$, all integrations are finite without cutoffs.
At $d=3$, the integration involving the soft
mode (diagram ${\cal L}_2$) is
infra-red divergent.
Let us concentrate on the singular
piece of Eq.\,(\ref{eqn_of_state}) as $d\rightarrow 3$.
Integrating over $q'$ first gives:
\begin{eqnarray}
&&\int_{BZ} d^2 k'\int d^{d-2} q'
\frac{1}{\epm(k')/\xic^2+q'^2}
\nonumber\\
&\sim&
\int_{BZ}d^2 k'[ \epm(k')]^{\frac{d-4}{2}}
\Gamma\left(\frac{d-2}{2}\right)\Gamma\left(\frac{4-d}{2}\right)
\nonumber\\
&\sim&   \int^\Lambda_0 k'^{2d-7}dk'.
\label{eqn_infra_red}
\end{eqnarray}

In the last step, a circular BZ of radius $\Lambda$ is assumed.
The integral becomes logarithmically divergent at $d=3$
indicating that $d_{LC}=3$.
Therefore the Abrikosov phase is unstable against
long-wavelength fluctuations in the thermodynamic limit.
Such a conclusion has also been reached using a similar analysis within
the harmonic
approximation \cite{Maki71,Moore89}.
The identification of $\epm(k)\approx c_{66}\ell^4k^4/2\alpha_0^2\ah$
for small $k$ points to the
nature of the soft mode which is responsible for
the destruction of ODLRO at $d=3$: it
is a long-wavelength elastic shear wave.
The hard mode, on the other hand, is
associated with the compressional mode of the lattice\cite{Moore89}.
The low energy excitations
about the
ground state, i.e. the soft mode,
can be described by an effective Hamiltonian
\cite{Moore89}:
\begin{eqnarray}
{\cal F}_{eff}& =& {\cal F}_{MF}
+\frac{1}{2} \sum_{q,k\in {\rm HC}}\left(
c_{66}\ell^4 k^4+\rho_s q^2 \right)\frac{|a_p|^2}{\alpha_0^2},
\label{eqn_soft_mode_functional}
\end{eqnarray}
where $\rm HC$ denotes half of the two-dimensional
first BZ. $c_{66}$ and
$\rho_s$
are the elastic shear modulus and
superfluid density respectively. The variable $a_p$ is just the Fourier
component of the phase {\em change} $\theta$ of the fluctuating order parameter
\begin{eqnarray}
\theta=\frac{1}{\alpha_0\sqrt{2V}}\sum_{p} a_{p} e^{i(k.r+q.{z})}.
\label{eqn_phase}
\end{eqnarray}
(Note that the constraint $a_p^*=a_{-p}$ in Eq.\,(\ref{eqn_phase})
guarantees that $\theta$ is real.)
The soft mode displacements ${\bf u}$ of the vortex
lines can be expressed
in terms of the derivatives of $\theta$:
$u_x=-\ell^2 \dif \theta/\dif y$ and $u_y=\ell^2 \dif \theta/\dif x$
\cite{Moore89}.
Since $\nabla.{\bf u}=0$,
the flux line motions associated with the soft mode
are shear waves\cite{Moore89}.
On introducing the dimensionless transverse and longitudinal
lengths ${\bf R}=(x/\ell,y/\ell)$
and ${\bf Z}={\bf z}/\xic$ respectively,
Eq.\,(\ref{eqn_soft_mode_functional}) becomes in these dimensionless
variables
\begin{eqnarray}
&&~\frac{{\cal F}_{eff}}{k_BT}=\frac{{\cal F}_{MF}}{k_B T}
\nonumber \\
&&+
\frac{1}{2\tT}\int d^2 {\bf R}
\int d^{d-2}{\bf Z} \left[
\tilde{\rho}_s\left(\frac{\partial \theta}{\partial {\bf Z}}\right)^2
+\tilde{c}_{66}\left(\nabla^2_\perp \theta\right)^2
\right],
\label{eqn_dimensionless_functional}
\end{eqnarray}
where $\tilde{c}_{66}=\betak c_{66}/2\ah^2$ and
$\tilde{\rho}_s=\mc\betak \rho_s/\hbar^2\ah$ are the dimensionless shear modulus and
superfluid density respectively. At the LLL mean-field level,
$\tilde{c}_{66}=0.0885\ldots$ and $\tilde{\rho}_s=1/\beta_A=0.8624\ldots$
respectively\cite{Moore97}.

The length scales $\ellc$ and $\ellab$
over which the ODLRO decays can be extracted from the
singular piece in Eq.\,(\ref{eqn_of_state}).
As mentioned before, we expect that $\ellab$ is also a measure
of the range of the crystalline order in the
transverse plane. It will be determined by
setting $\ar=0$ on the left hand side of Eq.\,(\ref{eqn_of_state})
and integrating $q \in (0,\infty)$ and then
$k\in \sqrt{2}(\ellab^{-1}, \ell^{-1})$.
We obtain
\begin{eqnarray}
\ellab\approx \ell \exp({\cal A}|\alpha_T|^{3/2}/2),
\label{eqn_ellab}
\end{eqnarray}
with ${\cal A}=2\sqrt{\tilde{c}_{66}\tilde{\rho}_s}=0.553\ldots$
Similarly, one can extract $\ellc$ by first integrating over
$k \in \sqrt{2} (0,\ell^{-1})$ and then $q$ from $2\pi(\ellc^{-1},\xic^{-1})$.
This yields
\begin{eqnarray}
\ellc\approx \xic \exp\left( {\cal A} |\alpha_T|^{3/2}\right).
\label{eqn_ellc}
\end{eqnarray}
As the temperature drops, $\ellab$ and $\ellc$ grow rapidly but
only diverge in the zero temperature limit ($\alpha_T\rightarrow -\infty$)
where true ODLRO order exists.
The functional forms for $\ellab$ and $\ellc$ have been suggested by
one of us\cite{Moore97} using a simple renormalization group argument based on
the effective Hamiltonian
Eq.\,(\ref{eqn_dimensionless_functional}).
However, the number ${\cal A}$ is estimated here
for the first time\cite{Kienappel97b}. We acknowledge that our
estimate of ${\cal A}$ can only be regarded as an order of magnitude estimate.
For example, setting $(\ar/\alpha_0)^2$ equal to a constant$<$1 rather than
zero produces a different value of ${\cal A}$. However, we
find that the value of ${\cal A}$ quoted here gives an excellent prediction
for the position of the ``melting'' line in YBCO (see below).
Our calculation of the ``jumps'' in the entropy, magnetization etc.
is independent of our estimate of ${\cal A}$, as we express their magnitudes
in terms of the measured value of $\alpha_T^*$ at the ``melting'' line.
In Appendix\,\ref{appendix_shear}, we suggest that only a non-perturbative
approach incorporating the topological defects such as entanglements will
lead to a quantitative estimate of the coefficient {$\cal A$}.

For crystals of the shape normally used in the
studies of high temperature superconductors, $\ellc$ will grow to
the system dimension $L_z$ before $\ellab$ reaches the transverse dimension
$L_x$.
When this happens, there is
phase correlation along the $c$-axis, and the
system will then behave as if it were effectively two dimensional.
$\ellab$ is also growing exponentially, although
slower than $\ellc$, and it is expected to be
several orders of magnitude times the lattice spacing at the crossover
temperature when $\ellc\approx L_z$.
Therefore, when $\ellc\approx L_z$,
the system is a vortex liquid with quasi-long range order,
which explains the apparent Bragg-like peaks in neutron scattering
experiments\cite{Cubitt93,Lee95}. The apparent
formation of sharp Bragg-like
peaks from the rings (expected to be) seen in the
structure factor for the vortex liquid phase when the temperature is lowered
is usually attributed to the freezing of the vortex liquid to the vortex
crystal phase. However, a recent theoretical investigation\cite{Yeo97}
has found that such a transformation in
the structure factor can also take place entirely
within the vortex liquid phase in the presence of a weak
four-fold symmetric coupling to the underlying crystal.
In fact, the angular width of
the peaks $\delta\theta $ varies as
$\delta \theta \propto \ellab^{-1}$
for a given coupling to the underlying crystal.
This implies that the width of the peaks should shrink exponentially
rapidly when the temperature is lowered.

\begin{figure}[htb]
\narrowtext
\centerline{
\epsfxsize= 8.5cm 
\epsfbox{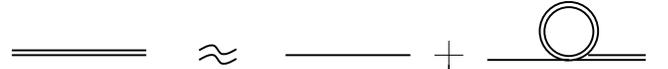}}
\vspace{0.5cm}
\caption{The Hartree-Fock approximation. The double and single
lines denote the
renormalized and bare propagators $G(q)$ and $g(q)$ respectively.} 
\label{fig_hf} 
\end{figure}

The mechanism for the 3D to 2D crossover can be illuminated
by a ``toy'' calculation for the 3D vortex liquid.
Consider the Hartree-Fock approximation to the propagator in the vortex liquid
phase given in Fig.\,\ref{fig_hf}. This calculation is normally
done for an infinite system, but we shall do for a system of finite width $L_z$
to illustrate the 3D-2D crossover mechanism.

The renormalised propagator $G(q)$
(deriving from $<\Psi^*(\br',{\bf z'})\Psi(\br,{\bf z})>$)
is related to the bare propagator $g(q)=k_BT/(\alpha_H+\hbar^2q^2/2\mc)$
such that the ``mass'' term $\alpha_H$ becomes $G(q)=k_BT/(\alpha_R+\hbar^2q^2/2\mc)$
with 
\begin{eqnarray}
\alpha_R &=&\alpha_H+\frac{\betak k_B T}{\pi\ell^2L_z}
\sum_{n=-\infty}^\infty\frac{1}{\alpha_R+(\hbar^2/2\mc)q_n^2}
\nonumber \\
&=& \alpha_H+\frac{\betak k_B T}{2\pi \ell^2}
\sqrt{\frac{2\mc}{\hbar^2 \alpha_R}}
\coth\left(\frac{L_z}{\tilde{\xi}_\parallel}\right)
\label{eqn_alphaR}
\end{eqnarray}
where the wavevector $q_n=2\pi n/L_z$ for a system of finite size $L_z$
with periodic boundary condition.
$\tilde{\xi}_\parallel=(\hbar^2/2\mc\alpha_R)^{1/2}$ is the
renormalized correlation length along the field direction.
By using the asymptotic behavior
$\coth(x)\sim 1/x$ for small $x$ and $\coth(x)\sim 1$ for large $x$,
one can see that Eq.\,(\ref{eqn_alphaR}) reduces to the well known results for
2D and 3D (Eqs.\,(23) and (24) in Ref.\,\onlinecite{Ruggeri76})
in the limit of $L_z<<\tilde{\xi}_\parallel$ and
$L_z>>\tilde{\xi}_\parallel $
respectively. Notice that the 2D limit ($L_z<<\tilde{\xi}_\parallel$)
of $\alpha_R$ is
dominated by just the $n=0$ term in the sums.
Although $\tilde{\xi}_\parallel$ is not exponentially growing
(because $d_{LC}=4$ in this approximation), one can see how
the behavior of the system is controlled by the ratio of the length scale
$L_z$ and $\tilde{\xi}_\parallel$, and that in the 2D limit one can
proceed as if the flux lines were straight rods.

Returning to the full problem, as the temperature change required to pass from the regime
$\ellc\ll L_z$ to $\ellc\approx L_z$ is very small\cite{Moore97},
we believe
that the crossover has been mistakenly
interpretated as a first order melting phase transition.
Later on, we will calculate the sharp step in the magnetization and the
entropy/vortex/layer due to this crossover and show that
it has many features of a first order phase transition.

First, we shall investigate the position in the $H-T$ phase diagram where
the crossover
$\ellc \approx L_z$ takes place.
In dimensionless units this occurs when $\alpha_T=\alpha_T^*$ where
\begin{eqnarray}
\alpha_T^*\approx -\left[\frac{1}{{\cal A}} \ln
\left(\frac{L_z}{\xic}\right)\right]^{2/3}.
\label{eqn_crossover}
\end{eqnarray}
For a typical YBCO crystal of thickness 0.2mm and $\xic\approx 10{\rm \AA}$,
we estimate using our estimated value of ${\cal A}$ that
the crossover is at around $\alpha_{T}^*\approx -7.9$ which agrees
with the supposed `melting' line in previous investigations
\cite{Wilkin93,Hikami91}.
For the same sample thickness, but using a typical shorter coherence
length $\xic\approx 2.0 {\rm \AA}$ appropriate to BSCCO,
we find that $\alpha_T^*\approx -8.9$.
Note that the position of the crossover $\alpha_T^*$
is only weakly (logarithmically) dependent on $\L_z/\xic$, whose
dependence on $T$ and $B$ will therefore be neglected
in what follows below.
The dependence of the parameter $\alpha_T^*$ on $T$ and $B$
implies that the position of the crossover line in the phase diagram
should follow the power law:
\begin{eqnarray}
B_{cr}\approx B_0 \left(1-\frac{T}{T_c}\right)^{n},
\label{eqn_power_law}
\end{eqnarray}
where $n=3/2$ and $B_0$ is the zero temperature melting
magnetic field. Strictly speaking $T_c$ is the mean-field
transition temperature,
and fluctuation effects will make its value slightly
different from the measured zero-field transition temperature.

\begin{figure}[htp] 
\narrowtext
\centerline{
\epsfxsize= 8.8cm 
\epsfbox{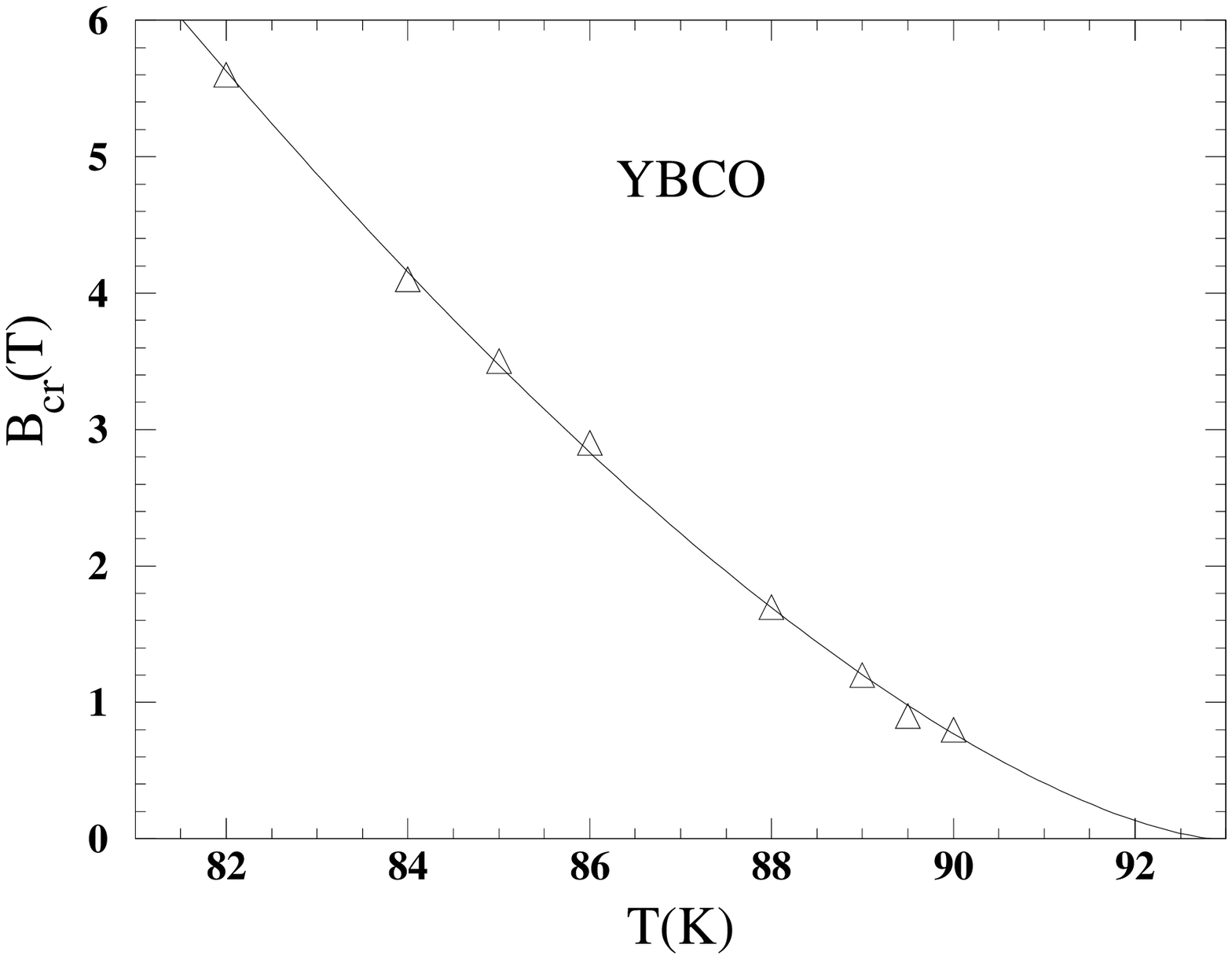}}
\vspace{-0.3cm}
\centerline{
\epsfxsize=9.0cm
\epsfbox{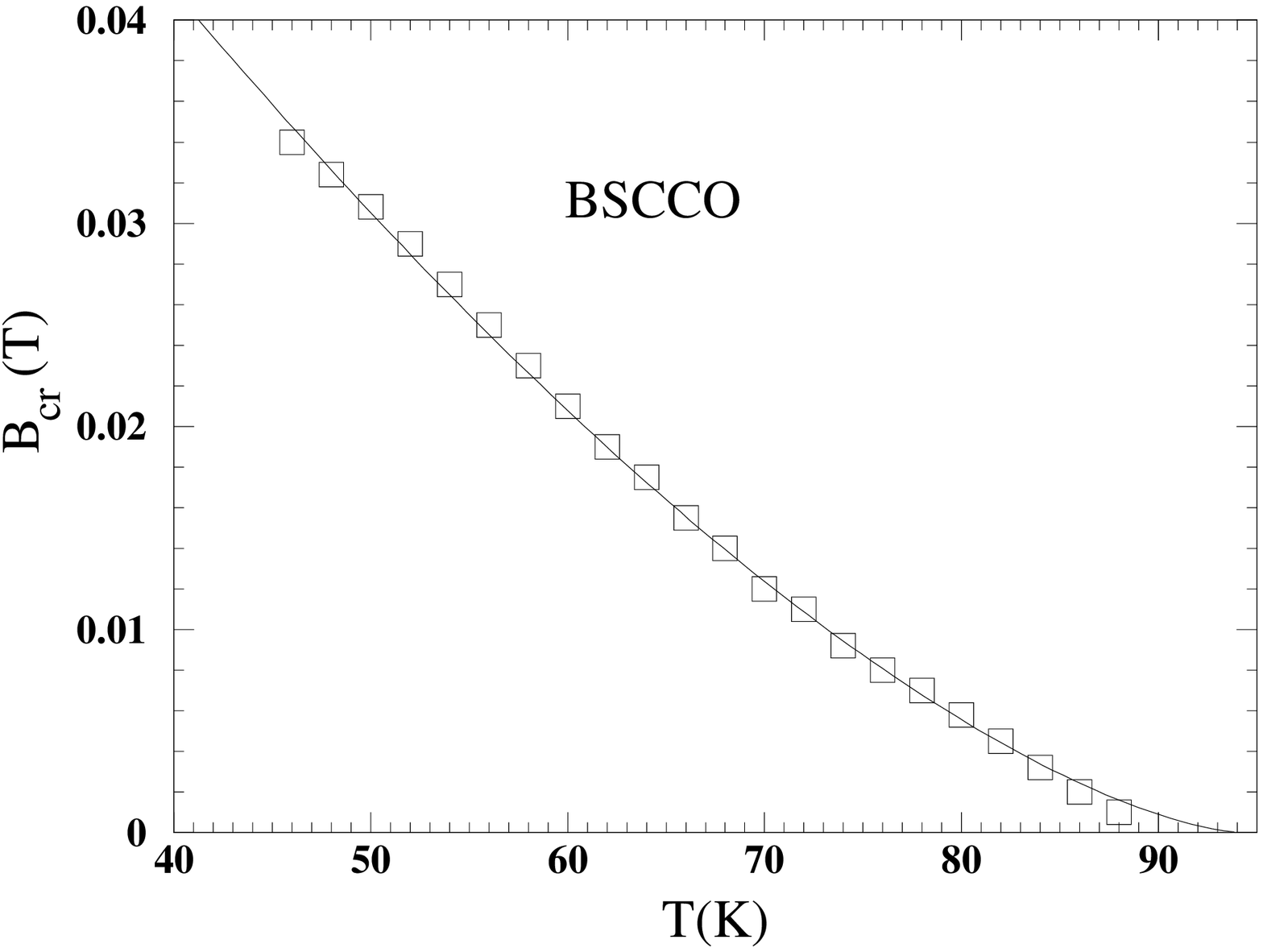}} 
\caption{The experimental `melting' points fitted with
Eq.\,(\protect \ref{eqn_power_law}) (solid line)
for YBCO and BSCCO. These data are read from the bottom panel
of Fig.\,5 in
Ref.\,\protect\onlinecite{Welp96} and
Fig.\,4 in Ref.\,\protect\onlinecite{Zeldov95} respectively. } 
\label{fig_melting} 
\end{figure}

Using the YBCO and BSCCO `melting' data from
Refs.\,\onlinecite{Welp96} and \onlinecite{Zeldov95}
respectively,
we find reasonably good fits with
$(B_0,T_{c})\approx$ (140T, 92.9K) and $(B_0,T_{c})
\approx$ (0.1T, 94.3K ) respectively (see Fig.\,\ref{fig_melting}).
By substituting Eq.\,(\ref{eqn_power_law}) into the definition of
$\alpha_T$, we have
\begin{eqnarray}
\alpha_T^*\approx -\left(\frac{2}{Gi}\right)^{1/3}
\left(\frac{B_{c2}(0)}{B_0}\right)^{2/3}.
\label{eqn_melting_line}
\end{eqnarray}

We can then determine the Ginzburg number $Gi$ of YBCO and BSCCO
by comparing
Eq.\,(\ref{eqn_crossover}) and Eq.\,(\ref{eqn_melting_line}).
Using $B_{c2}(0)\approx 170 $T in both cases \cite{bc2},
we obtain $Gi_{{\rm YBCO}}\approx 0.006$ and $ Gi_{\rm BSCCO}\approx 8000$.
While the former is a widely quoted number for YBCO,
the later is about four order of magnitude bigger than
the usual value quoted of 0.1.
We believe that this might not be unphysical for BSCCO,
and the argument is as follows.
What we are using is a phenomenological
model---the LLL approximation---which is quantitatively useful
as long as the effective temperature $\alpha_T$ provides a good
representation
of the true temperature and field dependence.
This seems to be the case for YBCO.
On the other
hand, $B_0$ for BSCCO is approximately three orders of magnitude
smaller than that of YBCO which suggests that fluctuations effects are enormous
and are likely to renormalize the bare parameters of the theory.
Effects from the higher Landau level contributions, the
quasi-two dimensional behavior of BSCCO
and the fluctuations of the vector potential,
which have been neglected in this effective model
will all act to modify the dependence
of $\alpha_T$ on temperature and magnetic field away from the
simple bare expression for $\alpha_T$ as in Eq.\,(\ref{eqn_melting_line}).
So in principle, one also should expect 
$Gi_{{\rm BSCCO}}$ to be a function of both temperature and magnetic field
rather than a constant, as we have done above.
However, a qualitatively
useful description of BSCCO may be possible
if we are prepared to accept a {\em renormalized} $Gi$ much larger
than the bare $Gi$.
Therefore, we will keep an open mind, and proceed to
investigate what the effective model can offer in the description
of both YBCO and BSCCO.
This philosophy seems to have been adopted by other
authors as well. In their recent Monte
Carlo simulation, Hu and MacDonald\cite{Hu97} had implicitly used a very
large $Gi$ in order to fit their numerical results to BSCCO.
On the other hand it is possible that the large value of $Gi$
needed to fit the data in BSCCO is really telling us that
the mechanism of the two transitions in YBCO and BSCCO are quite different
e.g. that the crossover idea applies to YBCO but that there is a genuine
first order transition in BSCCO.

Also of some interest is the angular dependence of the crossover line.
When the magnetic field is tilted at an angle $\phi$ to the $c$-axis,
then the crossover effect will occur at $\ellc \approx L_z/\cos(\phi)$.
Using the general scaling approach of Blatter \et\cite{Blatter92},
the dimensionless temperature scales as
(ignoring as before logarithmic corrections)
\begin{eqnarray}
|\alpha_T^*(\phi)|^{3/2}& \approx &|\alpha_T^*(0)|^{3/2}
\sqrt{\cos^2\phi+\sin^2\phi/\Gamma^2}
\end{eqnarray}
where $\Gamma=\sqrt{\mc/\mab}$ is the anisotropy factor.
Throughout
the rest of the paper, we will only examine the case of $\phi=0$,
i.e. a field along the $c$-axis.

\section{Thermodynamics}
\label{sec_thermodynamics}

Although the crossover behavior discussed above is a finite size
induced effect rather than a phase transition,
we can still associate with it
a ``jump'' in the entropy per vortex per layer
$\Delta S_{cr}$ and the magnetization $\Delta M_{cr}$
due to the narrowness of the crossover region.
We believe that apparent first order melting
signatures like the latent heat and magnetization
jumps observed in experiments are
in fact due to the entropy and magnetization
differences between the 2D and 3D vortex
liquid.

Although the jumps in the magnetization and entropy
are reported to be
sharp, they in fact have a finite width\cite{Welp96,Schilling96,Roulin96}.
Such rounding of the jumps is usually attributed to the
sample or magnetic field inhomogeneity. However, we can explain
such rounding as the natural width of the crossover effect.
A simple prescription to estimate the width is to find the small change
$\delta \alpha_T$ in $\alpha_T^*$ (set by $\ellc\approx L_z$)
required for $\ellc\approx 2L_z$.
Using the definition of $\alpha_T$, we obtain
$\delta \alpha_T/\alpha_T\approx (2/3) \delta B/B$. Substituting this
into Eq.\,(\ref{eqn_crossover}) and using $L_z=0.2$mm and
$\xic\approx 10{\rm \AA}$ for YBCO, we get $\delta B/B\approx 0.057$
at a magnetic field of $4.2$T.
Using the YBCO magnetization data from Fig.\,1 in Ref.\,\onlinecite{Welp96},
we estimate that the width is $\delta B\approx 0.2$T
at $B=4.2$. This gives $\delta B/B\approx 0.05$, which is
good agreement with the predicted crossover width.

If one can calculate the 2D and 3D free energies
$F_{2D}$ and $F_{3D}$ of the vortex liquid phase,
then one would naively expect the
$\Delta S_{cr}=-(s\Phi_0/BV)\partial(F_{3D}-F_{2D})/\partial T$
and $\Delta M_{cr}=-(1/V)\partial(F_{3D}-F_{2D})/\partial H$.
However, there are two subtleties involved here and this expectation
is not correct.
The first of these is a simplification.
From their definitions $\alpha_T$ and $\alpha_{2T}$ are such that
$\alpha_{2T}^*=-\sqrt{L_z|\alpha_{3T}^*|^{3/2}/4\xic}$.
Because $L_z\gg \xic$ for a bulk system, $\alpha_{2T}$
is orders of magnitude larger than $\alpha_{T}$.
For YBCO, the crossover occurs at
$\alpha_{3T}^*\approx -7.9$. For the 2D liquid, this corresponds
to $\alpha_{2T}^*\approx -1200$ (similar estimates apply to BSCCO).
At such a low effective temperature,
the behavior of the 2D liquid is basically
mean-field like, and fluctuation effects are negligible.
With this in mind, the sharp changes in the thermodynamic functions
between the two regimes can be obtained by
subtracting the mean-field expression
from that of the 3D expression.
The second subtlety is that not all contributions
to the entropy or magnetization are sensitive to the effect of crossover.
For example, the short wavelength contributions are not modified when
$\ellc$ becomes comparable to $L_z$, and
so will not contribute to the jumps. Therefore,
in calculating the 3D entropy, magnetization and specific heat jumps,
we need to examine all contributions and discard the pieces that
are continuous over the crossover region.

Before deriving the thermodynamic
functions, we would like to specify how we envisage infinitesimal
changes in the magnetic field, i.e. taking the derivatives of say, the
free energy with respect to the magnetic field.
We assume a finite system which is allowed to change its
transverse area $L_xL_y$ so that as the magnetic
induction changes, the number of vortices inside the system,
$N_\phi$ remain constant. The two are related by $N_\phi=L_xL_y/2\pi\ell^2$.
This framework naturally allows small changes in the
magnetic induction without the introduction of extra vortices,
and because of its calculational convenience
it has also been used in the Monte Carlo simulations
of vortices\cite{Sasik95b,Dodgson97,Kienappel97a}.

The total 3D free energy of the system can be written as
\begin{eqnarray}
F&=&F_{MF}-
N_\phi\frac{k_B T L_z\pi}{2\beta_A\xic\tT} {\cal G}(\tT) \nonumber\\
{\cal G}(\tT) &=&-{\cal E}^{(1)}\tT+{\cal E}^{(2)}\tT^2+\dots,
\label{eqn_free_energy}
\end{eqnarray}
where $F_{MF}$ is the mean-field free energy
and ${\cal G}(\tT)$ is a the 3D dimensionless free energy
calculated by expanding about the mean-field solution.
${\cal E}^{(i)}$ is a number
given by the
$i$-th loop contribution to the dimensionless free energy
(see Appendix \ref{appendix_free_energy}).
By definition, the total entropy per unit volume is

\beginwide
\begin{eqnarray}
S=-\frac{1}{V}\frac{\partial F}{\partial T}
=-\frac{1}{V}\frac{\partial F_{MF}}{\partial T}+
\frac{N_\phi\pi k_BL_z}{2\beta_A V}
\left[\frac{{\cal G}(\tT)}{\tT}\frac{\partial }{\partial T}
\left(\frac{T}{\xic} \right)+
\frac{T}{\xic}\frac{\partial}{\partial T}
\left(\frac{{\cal G}(\tT)}{\tT} \right)\right].
\end{eqnarray}
\endwide

The first term corresponds to the mean-field entropy per unit volume.
Across the crossover region, the first term inside the square brackets
is continuous because both ${\cal G}(\tT)$ and
$\dif(T/\xic)/\dif T$ are continuous.
It is the first derivative of ${\cal G}(\tT)$ in the second term
that gives the impression of a discontinuity
at the crossover to the 2D regime.
Therefore the apparent drop in entropy upon cooling through the
crossover region is the total entropy minus all the background pieces
(including the mean-field contribution),
that are continuous
or smoothly varying in the crossover region.
Thus, the jump in the entropy per unit volume
due to the crossover effect is:
\begin{eqnarray}
\Delta S=\frac{N_\phi L_z k_B T\pi}
{2\beta_A V\xic} \frac{\partial }{\partial T}
\left[\frac{{\cal G}(\tT)}{\tT} \right].
\end{eqnarray}
Per vortex per layer, the leading term in the loop expansion for
the crossover entropy jump is
\begin{eqnarray}
\Delta S_{cr} &=&\frac{s\pi k_B {\cal E}^{(2)} }{2\beta_A\xic}
\frac{\partial \tT}{\partial T}
\nonumber\\
&\approx& \frac{3 \pi^2 s k_B {\cal E}^{(2)} }{\beta_A\xic(0)}
\left(\frac{2}{Gi}\right)^{1/6} \frac{t^{*2/3} h^{*-1/3}}{|\alpha_T^*|^2},
\label{eqn_entropy_jump}
\end{eqnarray}
where the superscript $*$ mean that the quantities are evaluated at
the crossover.
In the last line, we have used the approximation
$1/t^*\ll 3/2(1-h^*-t^*)$ in calculating $\partial \tT/\partial T$
at the crossover.
(This approximation can be easily justified by comparing the order
of magnitude of the two terms using typical YBCO and BSCCO parameters
at the crossover. In fact one can establish that
$1/t^* \ll 3/2(1-h^*-t^*) \ll 1/h^*$).

In the same way, we can obtain the magnetization of the system
via the definition $M=-(1/V)\partial F/\partial H$.
(This means that the magnetization
is not just---$<\overline{|\Psi|^2}>$
as in the original work of Abrikosov\cite{Abrikosov57}).
Using the same argument as before,
the relevant crossover magnetization jump
arises from the term
$\partial[{\cal G}(\tT)/\tT]/\partial H$.
The leading crossover magnetization jump is
\begin{eqnarray}
\Delta M_{cr} &=&\frac{\pi \mu_0 H k_B t
{\cal E}^{(2)}}{2\beta_A \Phi_0\xic}
\frac{\partial \tT}{\partial H}
\nonumber \\
&\approx& \frac{2\pi^2 k_BT_c\mu_0 {\cal E}^{(2)} t^{*4/3} h^{*1/3}}
{\beta_A\xic(0)\Phi_0|\alpha_T^*|}
\left(\frac{Gi}{2}\right)^{1/6}.
\label{eqn_magnetization_jump}
\end{eqnarray}
Again, we have used the approximation $1/h^*\gg 3/2(1-h^*-t^*)$
in calculating $\partial \tT/\partial H$ at the crossover.

It is now easy to see that $\Delta S_{cr}$ and $\Delta M_{cr}$
satisfy the Clausius-Clapeyron Eq.\,(\ref{eqn_claperon})
automatically.
The gradient of the crossover
$\partial H_{cr}/\partial T$ can determined by
differentiating Eq.\,(\ref{eqn_power_law}) directly.
This apparent thermodynamic
consistency in the jumps $\Delta S_{cr}$ and $\Delta M_{cr}$
has been used to
argue for the existence of first order melting in
YBCO\cite{Liang96,Welp96,Schilling96} and
BSCCO\cite{Zeldov95}.
However, our crossover scenario seems to provide a
possible alternative explanation.

\begin{figure}
\narrowtext
\centerline{\epsfxsize= 9cm 
\epsfbox{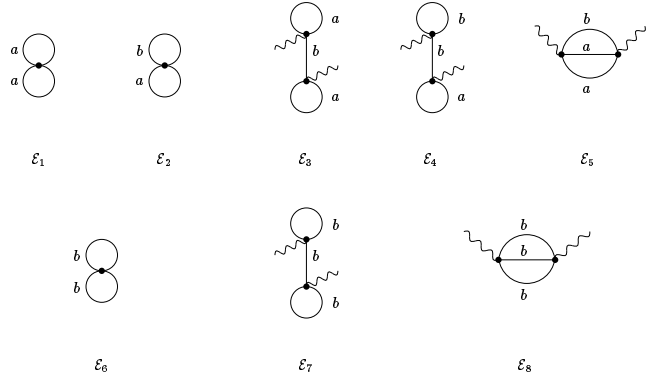}}
\caption{Free energy diagrams of ${\cal O}(\beta_k)$.} 
\label{fig_2loop_free_energy}
\end{figure}

How do our expressions for $\Delta M_{cr}$ and
$\Delta S_{cr}$ compare with the actual experimental results?
Notice that it is the two loop term ${\cal E}^{(2)}$ in the free energy,
rather than
one loop term ${\cal E}^{(1)}$ which gives
the leading order contribution to $\Delta S_{cr}$ and
$\Delta M_{cr}$.
In order to get an estimate of
$\Delta S_{cr}$ and $\Delta M_{cr}$ we need to calculate ${\cal E}^{(2)}$,
i.e. evaluate the diagrams shown in Fig.\,\ref{fig_2loop_free_energy}.

Each vertex is ${\cal O}(\betak)$, and each contribution from wavy line
is proportional to $\betak^{-1/2}$, and so all the
above diagrams are of the same order.
The single and double vertex diagrams have an overall negative and positive
sign respectively.
$a$ labels the soft mode propagator, which is infra-red divergent
at $d=3$. The sum of {\em all}
diagrams involving the soft mode remains finite although individual
diagrams are divergent
(see Appendix \ref{appendix_free_energy}).
The number ${\cal E}^{(2)}$ is estimated to be
$4.4\times 10^{-2}$.

Having found ${\cal E}^{(2)}$,
we can evaluate the orders of magnitude
of $\Delta S_{cr}$ and $\Delta M_{cr}$ at some typical field.
These result
should apply to both YBCO and BSCCO provided that
the appropriate {\em phenomenological} parameters are used to
used to model them realistically.
For YBCO, we choose $s\approx 10{\rm \AA}$, $\xic(0)\approx 2.2{\rm \AA}$ and
$\alpha_{T}^*\approx -7.9$. This give us
$\Delta S_{cr}\approx 0.7 k_B$/layer/vortex
and $\Delta M_{cr} \approx 4.0\times 10^{-5} T $ at $4$T.
For BSCCO, we use $s \approx 11 {\rm \AA}$, $\xic(0)\approx 1.8{\rm \AA}$
and $\alpha_T^*\approx -8.9$, and we estimate that
$\Delta S_{cr}\approx  1.0 k_B$/layer/vortex and
$\Delta M_{cr} \approx 0.5\times 10^{-4} T $ at $5\times 10^{-3}$T.
These results are in good agreement with
experiment\cite{Zeldov95,Welp96,Schilling96}.

\begin{figure}[htp]
\narrowtext
\centerline{\epsfxsize= 9cm 
\epsfbox{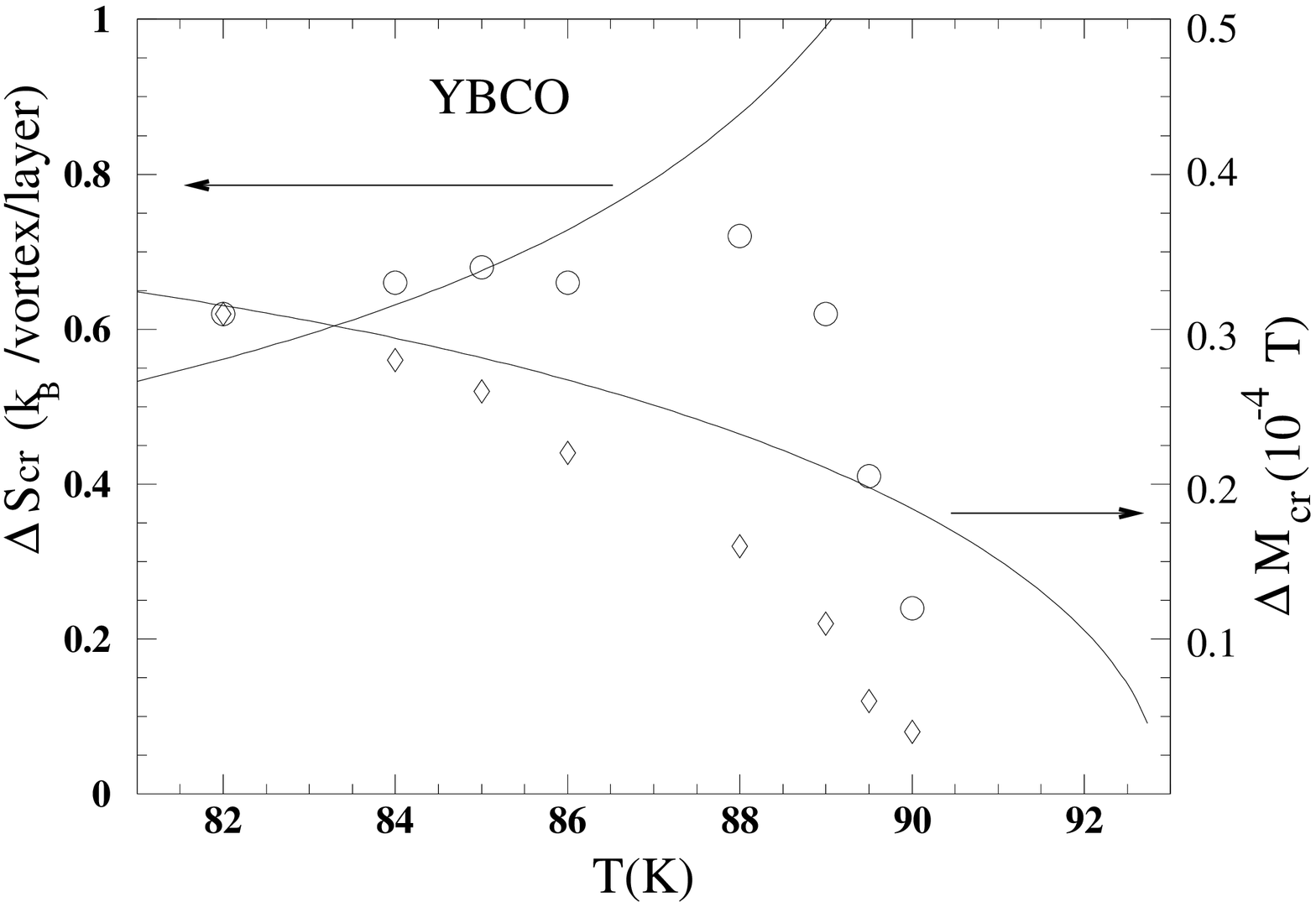}}
\centerline{\epsfxsize= 9cm
\epsfbox{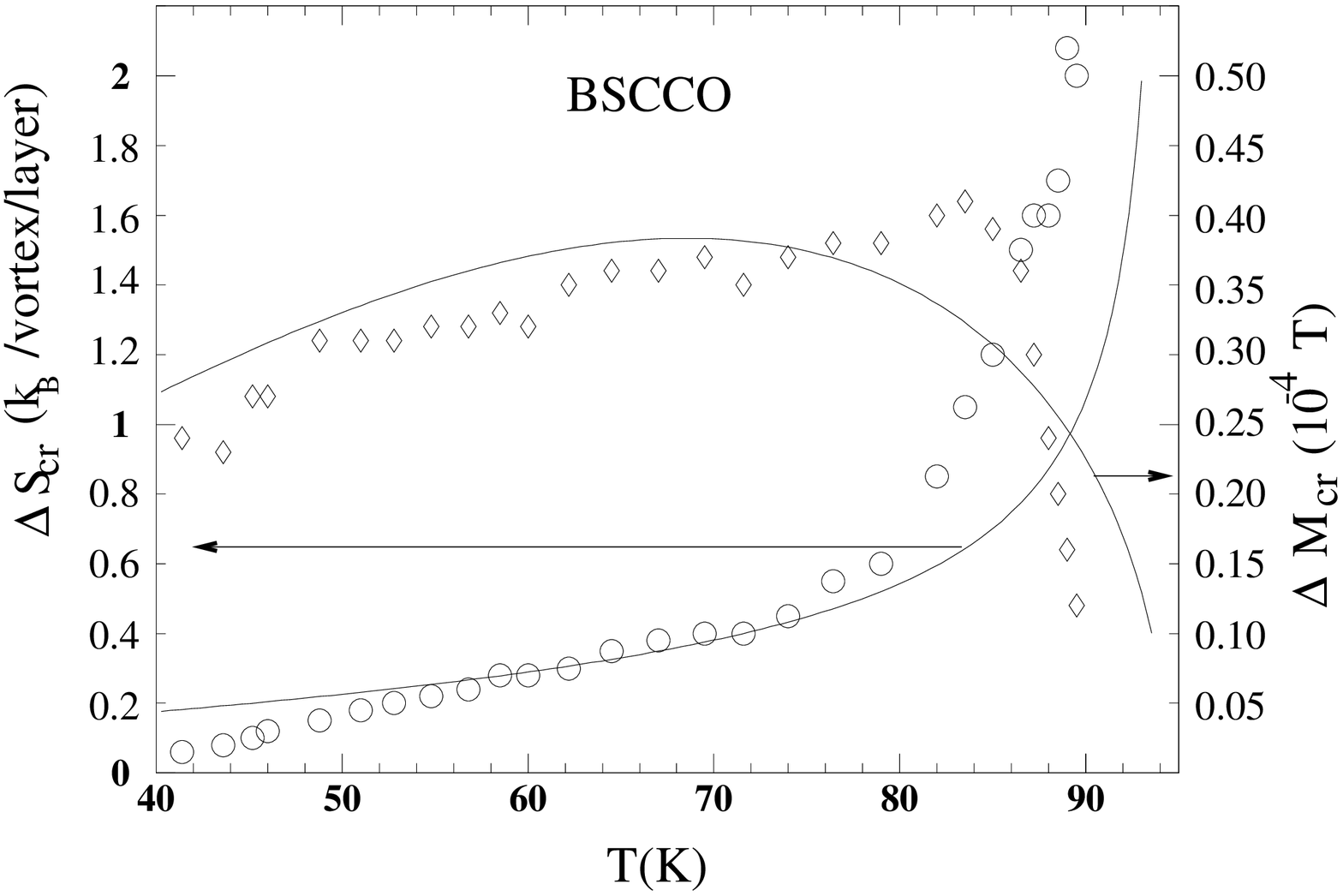}} 
\caption{$\Delta S_{cr}$ and $\Delta M_{cr}$ as a function of $T$
for YBCO and BSCCO. Data points for $\Delta S_{cr}$($\circ$) and
$\Delta M_{cr}$($\diamond$) are read from Fig.\,4(a) in
Ref.\,\protect\onlinecite{Welp96} and Fig.\,5 and 6 in Ref.\,\protect
\onlinecite{Zeldov95} respectively. Solid lines are theoretical fits
using Eq.\,(\protect\ref{eqn_magnetization_jump}) and
Eq.\,(\protect\ref{eqn_entropy_jump})
respectively, with material parameters discussed in
the text.} 
\label{fig_entropy} 
\end{figure}

More importantly, our crossover results predict that
$\Delta S_{cr} \sim h^{*-1/3}\sim (1-t^*)^{-1/2}$
and $\Delta M_{cr} \sim h^{*1/3}\sim (1-t^*)^{1/2}$ as $T$ approaches $T_c$.
Fig.\,\ref{fig_entropy}
shows the temperature dependence of $\Delta M_{cr}$ and
$\Delta S_{cr}$ compared to the
experiments using material parameters as previously mentioned.
The YBCO and BSCCO data points are
read from Welp \et\cite{Welp96}
and Zeldov \et\cite{Zeldov95}
respectively. 
Our result for $\Delta M_{cr}$ not only
agrees reasonably well 
with the bulk of the YBCO and BSCCO data, it also accounts for
the general temperature dependence quite well.
The divergence of $\Delta S_{cr}$ near $T_c$
is consistent
with the BSCCO data. However, this is apparently at odds with
the YBCO data, as Welp
\et \cite{Welp96} see vanishing $\Delta S_{cr}$
as $T \rightarrow T_c$. The authors themselves
suspect that this is due to the
influence of sample inhomogeneity very near $T_c$ \cite{Welp96}.
In view of this, more weight should perhaps be given to
the low temperature points when comparing our results
with the data in Fig.\,\ref{fig_entropy}.
We believe that if sample artifacts could be removed,
the divergence
of $\Delta S_{cr}$ as $T\rightarrow T_c$ in YBCO would be revealed.

A consistent picture of even the form of the crossover line is not
available from experiments.
Zeldov \et \cite{Zeldov95} deduced an exponent  $n\approx 1.55$ for BSCCO,
which is close
to the results expected from the crossover mechanism
(Eq.\,(\ref{eqn_power_law})).
On the other hand, Liang \et\cite{Liang96} and Welp \et\cite{Welp96}
deduced a YBCO
melting line with a smaller slope
($n\approx$ 1.34, 1.36 respectively).
The confusion is further exemplified by the YBCO
measurement of Nishizaki \et \cite{Nishizaki96} where
a power law appropriate to the London model ($n\approx 2$)
was used. They estimated that $\Delta S_{cr}\approx 6k_B$
at $1$T and $25 k_B$ at $3$T. Both the magnitude and the temperature
dependence of their results are in total disagreement with
our results here and other YBCO
measurements\cite{Welp96,Liang96,Schilling96}.

In our approach, there is no
qualitative difference between YBCO and BSCCO. They just differ in
having
vastly different {\em effective} values of $ Gi$.
But our treatment for BSCCO requires the insertion of a value of
$Gi$ hugely renormalized by fluctuation effects. $Gi$
would then be expected to be a function of temperature and field,
and not a constant as assumed here. Inserting
such temperature and field dependence (if known!) could improve the fits in
Fig.\,\ref{fig_entropy}.
Without inclusion of disorder, our model does not explain the vanishing of $\Delta S_{cr}$
at a lower temperature
critical point as observed in BSCCO by Zeldov \et\cite{Zeldov95}
(neither do the melting nor decoupling models).
Such behavior is thought to be disorder induced.
The effect of random disorder will be discussed in more detail
in Section\,\ref{sec_disorder}, and seems consistent with the data of
Ref. \onlinecite{Zeldov95}.

\begin{figure}[htp]
\narrowtext
\centerline{ \epsfxsize= 9cm
\epsfbox{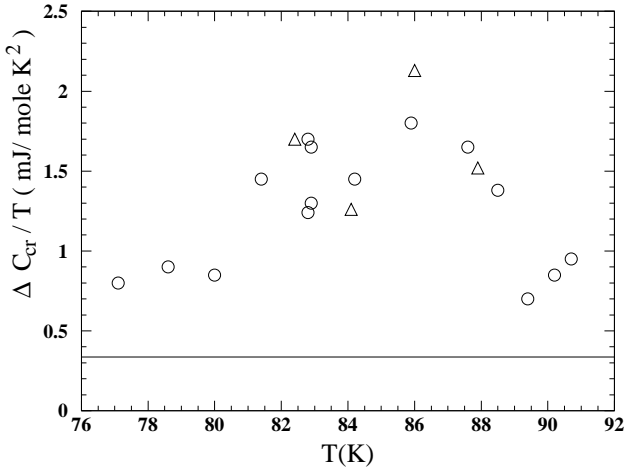} }
\caption{ Specific heat difference due to the crossover from
3D to 2D.
Data points with magnetic field parallel to the $c$-axis are read
from Fig.\,5 of Ref.\,\protect\onlinecite{Schilling97}.
The circular and triangular open symbols
represent $\Delta C_{cr}/T$ based on specific heat
and magnetization measurements
respectively. The solid horizontal line is our result using
Eq.\,(\protect \ref{eqn_deltaC/T_2}) using a set of parameters
appropriate to YBCO.}
\label{fig_heat} 
\end{figure}

The recent advent of reliable calorimetric
measurements\cite{Schilling97,Junod97} has also made available
specific heat data for comparison.
This has motivated us to calculate the leading crossover value
for the specific heat jump.
The total specific heat
capacity is given by
\beginwide
\begin{eqnarray}
\frac{C}{T}& = &\frac{\partial S}{\partial T}=
-\frac{1}{V}\frac{\partial^2 F_{MF}}{\partial T^2}
+\frac{N_\phi L_z k_B \pi}{2\beta_A V}
\left[
\frac{{\cal G} (\tT)}{\tT} \frac{\partial^2}{\partial T^2}
\left(\frac{T}{\xic}\right)+
\frac{T}{\xic}\frac{\partial^2}{\partial T^2}
\left(\frac{{\cal G} (\tT) }{\tT} \right)+
2\frac{\partial }{\partial T}\left(\frac{T}{\xic}\right)
\frac{\partial }{\partial T}\left(\frac{{\cal G} (\tT)}{\tT}\right)
\right].
\label{eqn_heat_capacity}
\end{eqnarray}
\endwide
The first term is just the mean-field part of the specific heat.
The first term inside the square bracket in Eq.\,(\ref{eqn_heat_capacity})
is continuous through the crossover region.
Only the last two terms with the derivatives
of ${\cal G}$ are sensitive to the crossover and hence
contribute to the crossover specific heat jump $\Delta C_{cr}$.
After some algebra, we find that to leading order in the loop expansion,
per vortex per layer,
\begin{eqnarray}
\frac{\Delta C_{cr}}{T}\approx \frac{9\pi^2{\cal E}^{(2)}s h^{-1}}
{2\beta_A\xic(0) T_c |\alpha_T|^3 }
\left(\frac{2}{Gi}\right)^{1/2}k_B.
\label{eqn_deltaC/T_1}
\end{eqnarray}
In order to make it more convenient to compare with the experiment
of Schilling \et \cite{Schilling97},
we convert this result to units of mJ/mole\,K$^2$ using the scale in
Fig.\,4(b) of Ref.\,\onlinecite{Schilling97} (namely, $1k_B$/vortex/layer
$\equiv$ 0.6 mJ/mole\,T\,K),
so Eq.\,(\ref{eqn_deltaC/T_1}) in the units of mJ/mole\,T\,K$^2$
is given by
\begin{eqnarray}
\frac{\Delta C_{cr}}{T}\approx 0.6\times \frac{9\pi^2{\cal E}^{(2)}s B_{c2}(0)}
{2\beta_A\xic(0) T_c |\alpha_T^*|^3 }
\left(\frac{2}{Gi}\right)^{1/2}.
\label{eqn_deltaC/T_2}
\end{eqnarray}
Using the appropriate parameters for YBCO, we found that
the leading $\Delta C_{cr}/T$ is constant at 0.3 mJ/mole\,K$^2$,
which is approximately four times smaller than the data in Fig.\,5
of Ref.\,\onlinecite{Schilling97} (see Fig.\,\ref{fig_heat}).
This discrepancy may be due to our neglect of
the higher order terms in the loop expansion.
In any case, as a first order approximation,
our results give a useful
qualitative description of the current experimental data.

\section{Effect of Random disorder}
\label{sec_disorder}

In this section, we investigate the effect of {\em quenched }
disorder (which is always present even in high quality
YBCO and BSCCO crystals), on the order in the Abrikosov phase.
The effects of quenched disorder on the ``melting'' transition
of YBCO and BSCCO at high magnetic fields is well documented.
We will just mention a few of them here.
Safar \et \cite{Safar93} reported the existence of
an ``upper critical point'' in untwinned YBCO
at a high magnetic field beyond which
the sharp drop in resistivity disappeared.
In a recent report of the
specific heat measurements on YBCO, Roulin \et \cite{Roulin96}
have suggested that
the termination of the ``melting line'' takes place at about 14T.
Using a sensitive local Hall probe,
Zeldov \et \cite{Zeldov95} also reported similar feature
in BSCCO, albeit at a much lower magnetic field (0.038T).
It is widely believed that at high magnetic fields the pinning of the vortices
by disorder is more effective
and the ``first order melting transition''
is removed\cite{Safar93,Fendrich95,Junod97}.
In an illuminating experiment,
Fendrich \et\cite{Fendrich95} have directly demonstrated that the suppression of
the sharp kink in the resistivity drop in YBCO is a disorder-induced effect.
They measured and compared the resistivity before and after a controlled
introduction, by electron irradiation,
of point defects in the sample. Furthermore, they showed that
the sharp kinks in the resistivity drop can be recovered by reducing the
density of the point defects through subsequent annealing of the sample
\cite{Fendrich95}.
At a low magnetic field, disorder is also thought
to affect the properties
of YBCO. The existence of
such ``lower critical point'' produced by disorder
or sample inhomogeneity (by an unknown mechanism) was recently invoked to explain the
disappearance of the jumps in magnetization and entropy/vortex/layer
in YBCO experiments\cite{Welp96,Schilling96,Junod97} at low fields.

With this motivation we proceed to investigate
the effect of disorder on the
Abrikosov phase.
For simplicity, we will neglect the effect of thermal fluctuations
and focus on very weak spatially varying
disorder characterized by a locally varying transition temperature
$\tau(\br)$. As a first approximation, we adopt the
``random $T_c$'' approach\cite{Ma_book,Larkin70}, i.e.
$\tau(\br)$ is assumed to have a Gaussian
distribution:
\begin{eqnarray}
\ll\tau(\br)\gg &=&0,
\nonumber\\
\ll\tau(\br)\tau(\br')\gg &=&\frac{W}{2}\delta (\br-\br'),
\end{eqnarray}
where $\ll\cdots\gg$ denotes averaging over all $\tau(\br)$
configurations.
The GL functional in the LLL approximation is given by 
\begin{eqnarray}
{\cal F}=\int d^d r\bigg\{ &&
\left(\alpha_H+\tau(\br)\right)|\Psi|^2
\frac{\hbar^2}{2\mc}\bigg|\frac{\partial \Psi}{\partial {\bf z}}\bigg|^2
\nonumber\\
&&+\frac{1}{2}\beta|\Psi|^4
\bigg\}.
\label{eqn_gl_disorder}
\end{eqnarray}
Our plan is to take the disorder to be weak and investigate
the effect of the disorder on the pure system ground state
$\Psi_0=\alpha_0\varphi(\br|0)$.
Expanding in terms of $a_p$ and $b_p$ about this,
we obtain the functional up to quadratic order

\beginwide
\begin{eqnarray}
{\cal F}& =&{\cal F}_{MF}+\frac{1}{2}\sum_p\ah\left[
(\epm+\xic^2 q^2)|a_p|^2+(\epp+\xic^2 q^2)|b_p|^2\right]
\nonumber\\
& &+\frac{2\alpha_0}{\sqrt{V}} \sum_p \int d^d r\left\{
\tau(\br)Re\left[
\left(ia_p+b_p\right)Q_k \varphi^*(\br|0)\varphi(\br|\br_k)
e^{iq.z}\right]\right\},
\label{eqn_gl_disorder_expansion}
\end{eqnarray}
\endwide
where ${\cal F}_{MF}$ is the ground state free energy.
The variables $a_p$ and $b_p$ are now disorder, rather than thermal,
induced fluctuation amplitudes
in the soft and hard modes respectively.
At this order, the minimum free energy is determined by the
conditions 
\begin{eqnarray}
\frac{\dif{\cal F}}{\dif x_p}=0,
\end{eqnarray}
where $x_p\in \{Re[a_p],Im[a_p],Re[b_p],Im[b_p]\}$.
A measure of the effect of the importance of the disordering effect
is $\delta$,
\begin{eqnarray}
\delta &\equiv&\frac{1}{\ar^2}\sum_{q,k\in{\rm HC}}\ll |a_p|^2\gg
\nonumber \\
&=&W\beta_A\sum_{q,k\in {\rm HC}}\frac{I(0,k|0,k)-|I(0,0|k,-k)|}
{(2\ah\epm+\gamma q^2)^2}.
\label{eqn_disorder_amplitude}
\label{eqn_delta1}
\end{eqnarray}
If $\delta$ is infinite it means that the disorder has perturbed the
system so much that the nature of the low-temperature state
is completely altered by it.
Integrating
over $q\in (0,\infty)$ in Eq.\,(\ref{eqn_disorder_amplitude})
first, and concentrating in the
long-wavelength fluctuations (small $k$), we obtain
\begin{eqnarray}
\delta\approx \frac{\eta \tilde{c}_{66}^{d/2-3}\beta_A^{d/2-1}
\Gamma(\frac{6-d}{2})}{ 2^{2d-3}\pi^{3d/2-4}\sqrt{3}}
\tilde{W}
\int^\Lambda_\epsilon d\tk \tk^{2d-9},
\label{eqn_delta2}
\end{eqnarray}
where $\tilde{W}=W/(2\ah^2\xic^{d-2}\ell^2)$
is a convenient measure of the strength of the disorder
analogous to the dimensionless temperature $\tT$ defined earlier.
Notice that the integral for $\delta$ is infra-red divergent below
$d=4$ as $\epsilon\rightarrow 0$, implying that
weak disorder will destroy the Abrikosov phase at and below four dimensions.
One can easily
show that a similar expression like Eq.\,(\ref{eqn_delta1})
for the hard mode is finite, for $d<4$.

Since the lower critical dimension $d_{LC}$ is four in the
presence of disorder,
it is expected that the crystalline
order in the three dimensional vortex liquid
will have a power law dependence on the
strength of the disorder.
What then is the range of this crystalline order $ R_\perp$
as a function of $\tilde{W}$?
The ordered phase will only exist if $\delta$ is small.
By setting $\delta\approx 1$, and inserting a lower cutoff in the
integral in Eq.\,(\ref{eqn_delta2}) corresponding to the
smallest wavevector which can be associated with the
crystalline order in the vortex liquid
phase, i.e. $\epsilon=(2\pi)^{3/2}\ell/R_\perp$,
gives for $d=3$
\begin{eqnarray}
R_\perp \approx \ell \left(
\frac{2^9\pi^3\tilde{c}_{66}^{3/2}\beta_A^{1/2}}{\tilde{W}}
\right)^{1/2}.
\label{eqn_r_perp}
\end{eqnarray}

The range of the $c$-axis phase correlation $R_\parallel$
can be obtained likewise by evaluating the integral
on the right hand side of  Eq.\,(\ref{eqn_delta2})
over $\tk\in (0,\Lambda)$ first,
and then integrating over $q\in 2\pi(\xic^{-1},R_\parallel^{-1})$.
This gives
\begin{eqnarray}
R_\parallel\approx \xic \left(
\frac{2^5\pi\tilde{c}_{66}}{\tilde{W}}
\right).
\label{eqn_r_parallel}
\end{eqnarray}

As expected, both $R_\perp$ and $R_\parallel$ are growing
algebraically, rather than exponentially, as a function of $\tilde{W}$,
and are only infinite in the limit of zero $\tilde{W}$.
The procedure used to obtain $R_\perp$ and $R_\parallel$
are similar in spirit to that of the treatment of weak disorder
by Larkin\cite{Larkin70}.

In order to investigate in detail
how the disorder modifies the crossover
effects discussed earlier, one would have to take into account
both temperature and disorder induced fluctuations
in the perturbation expansion about the mean-field
solution, which is a complicated task.
However, we can get a qualitative idea
by substituting the crossover values $T^*$ and $B^*$ into
Eq.\,(\ref{eqn_r_perp}) and (\ref{eqn_r_parallel}).
This would be expected to give, to the leading order,
the effect of disorder on
the crossover.
At the crossover, we take $\ell^2 \propto 1/B_{cr}$, $\xic\propto \ah^{-1/2}$
and $\ah\propto (1-h^*-t^*)\approx B_{cr}^{2/3}$.
This gives the magnetic field dependence of the order
in the vortex liquid phase at the crossover as
\begin{eqnarray}
R_\perp &\propto& 1/(B_{cr} W)^{1/2},
\\
R_\parallel &\propto& 1/(WB_{cr}^{1/3}).
\label{eqn_broadening}
\end{eqnarray}
Therefore the
order decreases as one moves along the crossover line
from low to high magnetic field.
Our scenario of the effect of disorder is that when it is weak,
so that $R_\parallel>L_z$, it plays little role in the thermodynamic properties
of the superconductor. If $R_\parallel<L_z$, it takes the role
of $L_z$ in our previous crossover calculation. For strong disorder,
$R_\parallel$ may be so small that the sharpness associated with the
crossover is removed and the ``jumps'' disappear. This at least seems
to explain the existence of an ``upper critical field''.

\section{Conclusion}
\label{sec_conclusion}

We have shown that within the framework of the loop
expansion, the Abrikosov
phase is destroyed by thermal fluctuations
at and below three dimensions in the thermodynamic limit
and the only thermodynamic phase above $H_{c1}(T)$ is the
normal vortex liquid phase.
However the range of the ODLRO, which is characterised by
$\ellab$ and $\ellc$ in 3D,
is growing exponentially upon cooling and diverges only
in the zero temperature limit.
We calculated the growth of $\ellc$ and $\ellab$
within the loop expansion
and argue that the apparent sharp features seen in YBCO and BSCCO
specific heat and magnetization experiments
are actually
due to the crossing over of the fluctuation behavior
from 3D to 2D when $\ellc$ becomes comparable to the system thickness.
This is in contrast with the widely held belief that
there exists a genuine first order
vortex crystal to liquid melting transition
well below $H_{c2}(T)$ line.
We demonstrated that the entropy/vortex/layer $\Delta S_{cr}$ and
the magnetization jump $\Delta M_{cr}$ due to the
crossover satisfy the Clausius-Clapeyron equation
without invoking the presence of a first order phase transition.
We also show that
$\Delta S_{cr}$ and $\Delta M_{cr}$
can give a reasonable account of the magnitude and
general temperature dependence of
the jumps in YBCO and BSCCO.
The only free parameter, $Gi$, is obtained by fitting
the experimental ``melting'' line.
Our estimate of the jump in the specific heat
of YBCO is also of the same order of magnitude
as the most recent specific heat measurements of
Schilling \et \cite{Schilling97}. Finally, we have investigated
the effect of quenched short-range disorder on the
Abrikosov phase. The lower critical dimensions $d_{LC}$
is found to be four, implying that the
short range order in the 3D vortex liquid phase has a
power law growth as the strength of the disorder is reduced.
We also demonstrated that random disorder tends to
remove the sharp crossover effect
as the magnetic field is increased, providing a possible
explanation for the existence of the upper critical field.

\acknowledgements
SKC acknowledges the support of ORS and a Manchester Research Studentship.
We benefited
from many discussions with J. Yeo and S. Phillipson.

\endcol

\appendix 
\section{Shear Modulus and Superfluid Density}
\label{appendix_shear}
\begin{figure}[htb]
\centerline{\epsfxsize=12cm
\epsfbox{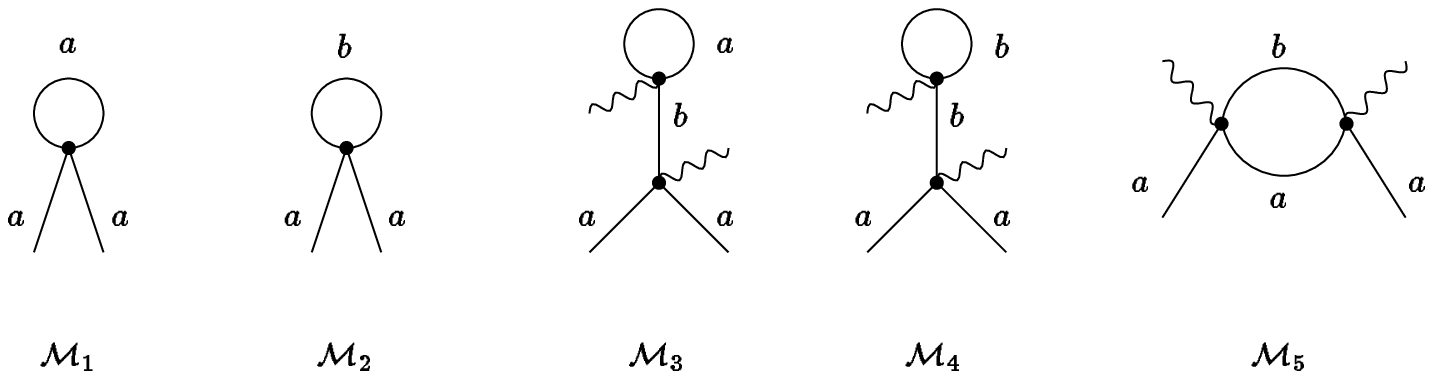}}
\caption{Two leg diagrams of order ${\cal O}(\betak)$.}
\label{fig_two_legs}
\end{figure}

In this section, we calculate the one loop correction to
the dimensionless superfluid density $\tilde{\rho}_s$ and
the elastic shear modulus $\tilde{c}_{66}$. We assume
that $d=3+\varepsilon$ where $\varepsilon$ is an arbitrarily
small number so that the Abrikosov lattice exists
at low temperature. The limit $\varepsilon \rightarrow 0$ will
then be taken to get results relevant to the bulk $d=3$ system.
Both $\tcsix$ and $\trhos$ to one-loop order
can be extracted from the equation of state
Eq.\,(\ref{eqn_of_state}) and the
renormalized soft mode propagator
$\tilde{G}_a^{-1}=G_a^{-1}+\sum_i {\cal M}_i+{\cal O}(\beta_\kappa^2)$,
which involves the two leg diagrams shown in Fig.\,\ref{fig_two_legs}.

Each of the above diagrams is of ${\cal O}(\beta_\kappa)$
since each vertex and wavy line contribute ${\cal O}(\beta_\kappa)$ and
${\cal O}(1/\sqrt{\beta_\kappa})$ respectively.
The full expression for the ${\cal M}_i$ and the calculation of $\rho_s$ is given in
Appendix \ref{appendix_ga}.
Writing $\tilde{\rho_s}= 1/\beta_A^{\Delta}-\tilde{\rho_s}^{(1)} \tilde{T}$,
we have
\begin{eqnarray}
\tilde{\rho_{s}}^{(1)} &= & \frac{\Gamma(\frac{4-d}{2})}{2^{d+1} \pi^{d/2+2}}
\int_{BZ} d^2 \tk \left(
\epp^{(d-2)/2}+\epm^{(d-2)/2}+\epp^{(d-4)/2}
+\frac{\epp^{(d-2)/2}-\epm^{(d-2)/2}}
{(d-2)|I(0,0|\tk,-\tk)|}\right).
\label{eqn_rhos_reduced}
\end{eqnarray}
For $d\rightarrow 3$,
Eq.\,(\ref{eqn_rhos_reduced}) can be estimated numerically to give
$\trhos^{(1)}\approx0.295$.

In principle, the one loop correction to $\tilde{c}_{66}$
can be extracted from the propagator Eq.\,(\ref{eqn_m})
in a similar manner by setting $q=0$ and expanding in $k$ about $k=0$.
However, it is more convenient to start from the definition of the elastic
shear modulus. Following Labusch \cite{Labusch69},
we allow the distortion of the ideal
triangular lattice with the constraint that the area of the primitive
cell of the first BZ is preserved.
$c_{66}$ is then defined as\cite{Labusch69}
\begin{eqnarray}
c_{66}=\frac{\eta^2}{V}\frac{\partial^2 F(\eta)}{\partial \eta^2}
\bigg\arrowvert_{\eta_\Delta},
\label{eqn_c66}
\end{eqnarray}
where $\eta$ is a dimensionless variable
specifying the shape of the unit cell with area $2\pi \ell^2$ (see Appendix
\ref{appendix_i}) and $F(\eta)$ is the free energy as a
function
of $\eta$ as in Eq.\,(\ref{eqn_free_energy}). However, only the one
loop free energy is needed here.
The second derivative of $F$ is evaluated at the
value of $\eta$ appropriate for a triangular lattice,
denoted by $\eta_{\Delta}(=\sqrt{3}/2)$.
The mean-field shear modulus has been obtained by
Labusch \cite{Labusch69}, who found $\tcsix^{(0)}\approx 0.0885$.
By extending this method to one loop order we have:
\begin{eqnarray}
\tilde{c}_{66} &=& \tilde{c}_{66}^{(0)}+\tilde{c}_{66}^{(1)}\tT\\
\tilde{c}_{66}^{(1)}&=&
\frac{\eta^2\Gamma(\frac{4-d}{2})}{2^{d+1} \pi^{d/2+2}(d-2)}
\frac{\partial^2 \sigma (\eta,d)}{\partial \eta^2}
\bigg\arrowvert_{\eta_\Delta},
\label{eqn_c66_reduced}
\end{eqnarray}
where
\begin{eqnarray}
\sigma (\eta,d) & =& \int_{BZ}
d^2 \tk \left[\epp^{(d-2)/2}+\epm^{(d-2)/2}\right].
\label{eqn_tau}
\end{eqnarray}

\begin{figure}[htp]
\centerline{\epsfxsize= 9cm
\epsfbox{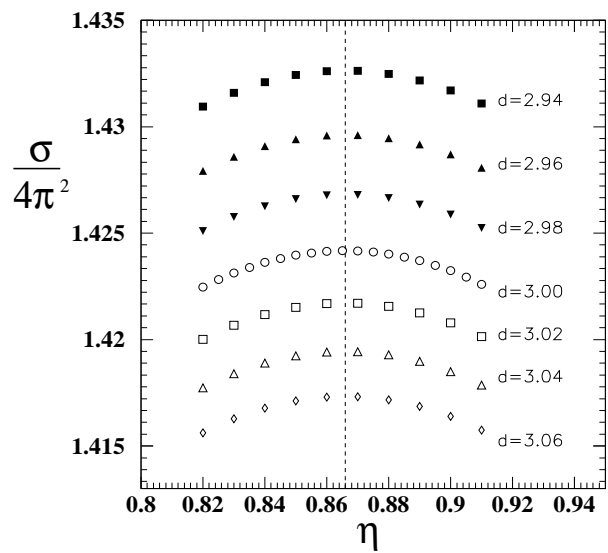}}
\caption{The function $\sigma/4\pi^2$ as a function of $\eta$ for different
$d$. The vertical dashed line denotes the position of the peaks at
$\eta_\Delta$.}
\label{fig_c66}
\end{figure}

Our results show that $\sigma(d,\eta)$ has a maximum at
$\eta_\Delta$
for any $d$ (see Fig.\,\ref{fig_c66}). This is consistent with the definition
Eq.\,(\ref{eqn_c66}) where the second derivative is evaluated at a saddle point,
i.e. at $\eta_\Delta$.
However, the {\it total} free energy would still have a minimum at $\eta_\Delta$
with the curvature of the free energy slightly altered
since the mean-field energy
$F_{MF}$ dominates for small $\tT$.
For $d\rightarrow 3$, we found using the data in
Fig.\,\ref{fig_c66} that $\tcsix^{(1)}\approx-0.0484$.

Because of the destruction of the Abrikosov lattice at $d=3$,
we might have expected that both $\tcsix$ and $\trhos$ should
be singular. However,
at one loop order both of these quantities are found
to be finite in the limit $d\rightarrow 3$.
The perturbative
one loop corrections to $\rho_s$ and $c_{66}$ are
not `aware' of the destruction
of the lattice at $d=3$ in the thermodynamic limit.

However, we believe that in a nonperturbative approach,
which would include topological defects such as entanglements,
both $c_{66}$ and $\rho_s$ would vanish.
Notice that in the $(2+\varepsilon)$ treatment of the O(n) model, $n\ge2$,
the mechanism of the transition is the vanishing of the superfluid
density, rather than the order parameter\cite{Binney_book}. In our case, 
neither $\rho_s$ nor $c_{66}$ is driven to zero by small amplitude
thermal fluctuations, and the mechanism of the transition is the vanishing
of ODLRO.

\section{The soft mode propagator and Equation of State}
\label{appendix_ga}
In this appendix, we give the explicit expressions for the one loop
diagrams discussed in Appendix\,\ref{appendix_shear} and in
Section\,\ref{sec_eqn_of_state}. Note that the expressions for ${\cal M}_5(p)$
are in abbreviated form, i.e. they are not explicitly {\it even} function of
the external momentum $p=(k, q)$. Such symmetry must be imposed.
The primed parameters $p'=(k',q')$ denotes
the internal momentum that has to be integrated over.
We shall also use the notation $\int d\vec{p'}=\int d^{d-2}q'd^2k'/(2\pi)^d$.
\begin{eqnarray}
{\cal L}_2 &=&
\frac{2\betak\beta_A\ar k_B T\sqrt{V} }{\gamma}\int d\vec{p'}
\left(\frac{\epm(k')+1}{\epm(k')/\xic^2+q'^2}\right)\\
{\cal L}_3&=& \frac{2\betak\beta_A\ar k_B T\sqrt{V} }{\gamma}\int d\vec{p'}
\left(\frac{\epp(k')+1}{\epp(k')/\xic^2+q'^2}\right)\\
{\cal M}_1(p)&=&\frac{-2\beta_\kappa\beta_A}{k_B T}\int d\vec{p'} G_{a}(p')
\left(I(k',k|k',k)+
2Re\left[Q_{k'}^2 Q_k^{*2} I(k,-k|k',-k')\right]\right)\\
{\cal M}_2(p)&=&\frac{-2\beta_\kappa\beta_A}{k_B T}\int d\vec{p'} G_b(p')
\left(I(k',k|k',k)-
2Re\left[Q_{k'}^2 Q_k^{*2} I(k,-k|k',-k')\right]\right)\\
{\cal M}_3(p)&=& \frac{\beta_\kappa\beta_A}{k_B T }\int d\vec{p'} G_a(p')
\bigg[\epm(k')+1\bigg]\bigg[\epm(k)+1\bigg]\\
{\cal M}_4(p)&= &\frac{\beta_\kappa\beta_A}{k_B T }\int d\vec{p'}G_b(p')
\bigg[\epp(k')+1\bigg]\bigg[\epm(k)+1\bigg]\\
\nonumber
{\cal M}_5 (p)&=& \frac{4\beta_\kappa|\alpha_H|\beta_A}{(k_B T )^2 }
\int d\vec{p'} G_a(p'-p) G_b(p') \times \\ \nonumber
& &
\Big\{\left|I(0,k'|k,k'-k)\right|^2+|I(0,k|k',k-k')|^2+
|I(0,k-k'|k,-k')|^2\\
\nonumber
\vspace{0.5cm}
& &~~~~~
+4Re\left[Q_{k'}^2 I(0,k|k',k-k')I(0,k-k'|k,-k')\right] \\
\nonumber
& &~~~~~
-4Re\left[Q_k^2 I(0,k'|k'-k,k)I(0,k'-k|k',-k)\right]\\
\nonumber
& &~~~~~
-4Re\left[Q_{k-k'}^2 I(0,k'|k'-k,k)I(0,k|k-k',k')\right]\\
\nonumber
\vspace{0.3cm}
& &~~~~~
+8Re\left[Q_{k'}^{*2}Q_k^{2}Q_{k'-k}^{2} I^2(0,k'|k,k'-k)\right]\\
\nonumber
& &~~~~~
+8Re\left[Q_{k'}^{2}Q_{k}^{*2}Q_{k-k'}^{2} I^2(0,k|k',k-k')\right]\\
\nonumber
& &~~~~~
+8Re\left[Q_{k'}^{2}Q_k^{2}Q_{k'-k}^{*2} I^2(0,k'-k|k',-k)\right]\\
\nonumber
\vspace{0.3cm}
& &~~~~~
+8Re\left[Q_k^{2}Q_{k'-k}^{*2} I(0,k-k'|k,-k')I(k-k',k'|0,k)\right]\\
\nonumber
& &~~~~~
-8Re\left[Q_{k'}^{*2}Q_{k}^{2} I(0,k'|k'-k,k)I(k-k',k'|0,k)\right]\\
& &~~~~~
-8Re\left[Q_{k'}^{*2}Q_{k'-k}^{2} I(0,k'|k,k'-k)I(k,-k'|0,k-k')\right]
\Big\}.
\vspace{0.5cm}
\label{eqn_m}
\end{eqnarray}

The renormalized soft mode propagator to one-loop order is such that
\begin{eqnarray}
\tilde{G}^{-1}_a(p)=G_a^{-1}(p)+{\cal M}_1(p)
+{\cal M}_2(p)+{\cal M}_3(p)+{\cal M}_4(p)+{\cal M}_5(p).
\label{eqn_ga}
\end{eqnarray}
The propagator $\tilde{G}_a(k)$
is still massless. This can easily shown by
setting the external momentum $p=0$.
Indeed, $\tilde{G}_a$
is expected to be massless to all orders in the loop expansion
\cite{Wallace_book}.
The superfluid density \cite{Martin65} can be calculated using
the following relation for small $q$:
\begin{eqnarray}
\rho_s q^2=2k_B T \tilde{G}^{-1}_a(q,k=0)
\tilde{\alpha}^2.
\end{eqnarray}
Substituting Eq.\,(\ref{eqn_of_state}) into Eq.\,(\ref{eqn_ga}), we have:
\begin{eqnarray}
\frac{\rho_s}{\alpha_0^2}&=&
\gamma-\frac{\beta_\kappa\beta_A k_B T}{\ah}\int d\vec{p'}\times \Bigg\{
\frac{\epp(k')+2}{\epp(k')/\xic^2+q'^2}
+\frac{4\ah|I(0,0|k',-k')|}{\gamma(\epp(k')/\xic^2+q'^2)^2}
\nonumber \\
& & ~~~~~~~~~~~~~~
+\frac{\epm(k')}{\epm(k')/\xic^2+q'^2}
+\frac{32\ah^2q'^2 |I(0,0|k',-k')|^2}
{\gamma^2(d-2)(\epm(k')/\xic^2+q'^2)(\epp(k')/\xic^2+q'^2)^3} \Bigg \}.
\label{eqn_rhos/appendix}
\end{eqnarray}
All integrals in Eq.\,(\ref{eqn_rhos/appendix})
are infrared convergent.
On integrating over
$q'$, Eq.\,(\ref{eqn_rhos/appendix}) yields Eq.\,(\ref{eqn_rhos_reduced}).
Similarly, one can calculate $c_{66}$ in principle via
\begin{eqnarray}
c_{66} \ell^4k^4=2k_B T \tilde{G}^{-1}_a(k\rightarrow 0,q=0)\tilde{\alpha}^2.
\end{eqnarray}
However, expressing ${\cal M}_i$ in terms of explicit polynomial of small $k$
is a challenging task. One would expected that the leading order
should be $k^4$ corresponding
to a dispersion appropriate to the elastic shear mode.
A more straightforward way of calculating the correction to $c_{66}$
is to use
the definition discussed in Appendix\,\ref{appendix_shear}.

\section{Loop expansion of the free energy}
\label{appendix_free_energy}
As discussed in Section \ref{sec_thermodynamics},
the starting point of our calculation of the jumps in the
magnetization and the entropy is the 3D free energy.
In this appendix, we derive
the 3D free energy expansion about the mean-field solution
to two loop order starting from the definition:
\begin{eqnarray}
F=-k_BT\ln\left(\prod_p
\int [ {\cal D}a_p] [ {\cal D}b_p]
\exp(-{\cal F}/k_B T)\right).
\label{eqn_partition_function}
\end{eqnarray}
At the one loop level, we substitute the Gaussian functional
Eq.\,(\ref{eqn_eigen}) into Eq.\,(\ref{eqn_partition_function}),
and we obtain the correction to the mean-field as:
\begin{eqnarray}
F^{(1)}& =&\frac{k_BT}{2}
\sum_{k,q}\left[\ln\left(\frac{2\ah\epp(k)+\gamma q^2}{2\pi k_B T}\right)
+\ln\left(\frac{2\ah\epm(k)+\gamma q^2}{2\pi k_B T}\right)\right]\\
&=& \frac{\ah^2V}{2\betak\beta_A}
\left[\frac{\beta_A\tT}{4\pi^3}\int d^2 \tk(\epp(\tk)^{(d-2)/2}+
\epm(\tk)^{(d-2)/2} )\right]
+ k_B T\sum_{k,q}
\left[\ln\left(\frac{\gamma q^2}{2\pi k_B T}\right)\right].
\label{eqn_first_loop_energy}
\end{eqnarray}
The first and second term in Eq.\,(\ref{eqn_first_loop_energy}) correspond to
$F^{(1)}-F^{(1)}(\alpha_H=0)$ and $F^{(1)}(\alpha_H=0)$ respectively.
The second term is dependent on an ultraviolet cutoff in
the longitudinal vector $q$. The cutoff is of the order of the
reciprocal of the layer spacing of the model.
Since we are interested in the long wavelength limit, we follow the
standard prescription of absorbing this term into the normal phase
energy (see Ref.\,\onlinecite{Ruggeri76}).
Using this results in Eq.\,(\ref{eqn_free_energy}),
with ${\cal E}^{(1)}\approx0.524$ for $d=3$.

For the sake of simplicity, we set $d=3$ at the outset when discussing
the two loop calculation.
The total two loop free energy can be written as
\begin{eqnarray}
{\cal E}^{(2)}=\frac{\beta_A^2}{8\pi^2} ({\cal E}_s+{\cal E}_h).
\end{eqnarray}
${\cal E}_s$ is the sum of energy diagrams involving the soft mode
and ${\cal E}_h$ is the sum of diagrams containing the hard mode only.
After integrating over the longitudinal
component $q$, and expressing the remaining transverse integral
in terms of the dimensionless BZ (to simplify notation, the tilde on $k$ and $k'$
are dropped), we have

\begin{eqnarray}
{\cal E}_s &=& \sum_{i=1}^{5}{\cal E}_i\\
{\cal E}_1 &=&
\int\int \frac{d^2 k d^2 k'}{(4\pi^2)^2}
\frac{-f_+(k,k') }{\sqrt{\epm(k)\epm(k')} }\\
{\cal E}_2 &=&
\int\int \frac{d^2 k d^2 k'}{(4\pi^2)^2}
\frac{-2f_-(k,k')}{\sqrt{\epm(k)\epp(k')} }\\
{\cal E}_3&=&
\int\int \frac{d^2 k d^2 k'}{(4\pi^2)^2}
\frac{(\epm(k')+1)(\epm(k)+1)}{2\sqrt{\epm(k)\epm(k')} }\\
{\cal E}_4&=&
\int\int \frac{d^2 k d^2 k'}{(4\pi^2)^2}
 \frac{(\epp(k')+1)(\epm(k)+1)}{\sqrt{\epp(k')\epm(k)} }\\
{\cal E}_5 &=&
\int\int\frac{d^2 k d^2 k'}{(4\pi^2)^2}
\frac{g(k,k')}
{\sqrt{\epp(k+k')\epm(k)\epm(k')}
(\sqrt{\epp(k+k')}+\sqrt{\epm(k)}+\sqrt{\epm(k'})},
\end{eqnarray}
where
\begin{eqnarray}
f_\pm(k,k')&=&I(k,k'|k,k')\pm 2Re\left[Q_{k'}^{*2}Q_{k}^2I(k,-k|k',-k')\right]\\
g(k,k')&=&16 [Re(Q_{k-k'}^*Q_{k'}Q_{k} I(0,k'-k|k',-k) )]^2 \nonumber \\
&+ &16 \{ Re[Q^*_kQ_{k'}Q_{k-k'} I(0,k|k',k-k')
+Q^*_{k'}Q_kQ_{k-k'}I(0,k'|k,k'-k)]\}^2   \nonumber \\
& -&32 Re\{Q_{k-k'}^*Q_kQ_{k'}I(0,k-k'|-k',k)\}
Re\{Q_{k}Q_{k'}Q_{k'-k} I(0,k|k',k-k')\}   \nonumber \\
& -&32 Re\{Q_{k-k'}^*Q_kQ_{k'}I(0,k-k'|-k',k)\}
Re\{Q_{k'}Q_kQ_{k'-k} I(0,k'|k,k'-k)\}.
\end{eqnarray}

Each integral involving $1/\sqrt{\epm}$ is
logarithmically divergent in the infrared-limit,
but the sum of the integrands of ${\cal E}_i$
is finite in the infra-red limit ($k',k\rightarrow 0$),
and hence ${\cal E}_s$ is finite. Numerically, we find that
${\cal E}_s\approx 0.8$.
The diagrams in ${\cal E}_h$ are given by the expressions:
\begin{eqnarray}
{\cal E}_h &=&\sum_{i=6}^8 {\cal E}_i\\
{\cal E}_6 &=&
\int \int \frac{d^2 k d^2 k'}{(4\pi^2)^2}
\frac{-f_+ (k',k)}{\sqrt{\epp(k)\epp(k')}}\approx -0.89\\
{\cal E}_7 &=&
\frac{1}{2}\left[\int \frac{d^2 k }{4\pi^2}
\left(\frac{1}{\sqrt{\epp(k)}}+\sqrt{\epp(k)}\right)\right]^2
\approx 2.07\\
{\cal E}_8 &=&\int \int \frac{d^2k d^2k'}{(4\pi^2)^2}\frac{h(k,k')}
{\sqrt{\epp(k'+k)\epp(k')\epp(k)}
[\sqrt{\epp(k+k')}+\sqrt{\epp(k')}+\sqrt{\epp(k)}]}
\approx 0.62,
\end{eqnarray}
where
\begin{eqnarray}
h(k,k') &=&~4 Re[Q_{k+k'}^2 I(0,k'|-k,k+k') I(0,k|-k',k+k')]\nonumber\\
  & &+8Re[Q_{k}^{*2}Q_{k'}^2 I(0,k|-k',k+k')I(-k,k'+k|0,k')] \nonumber \\
  & &+16 \left[Re(Q_{k'}^*Q_{k} Q_{k+k'}I(0,k'|-k,k+k'))\right]^2.
\end{eqnarray}
All integrals can be evaluated numerically, and we found that
${\cal E}_h\approx 1.80$.

\section{The first BZ and the Function $I(k_1,k_2\mid k_3,k_4)$}
\label{appendix_i}

In this appendix, we briefly outline a procedure
for evaluating integrals involving $\mbox{$I(k_1,k_2\mid k_3,k_4)$}$
over the first BZ.
Intgerals over the longitudinal vector $q$
can be done analytically, but
the integration
over $k$ in the first BZ is done numerically.

Restricting ourselves to the class of centered rectangular
lattices\cite{Kleiner64},
the fundamental unit cell of the vortex lattice
can be characterized by two primitive
vectors $\br_{\rm I}=(1,0)\ell_0$ and $\br_{\rm II}=(1/2,\eta)\ell_0$,
where $\ell_0$ is the spacing between vortices.
The flux quantization condition determines the area of the unit cell
as $2\pi \ell^2(=\eta\ell_0^2)$. 
The corresponding first BZ has an area of $4\pi^2/\eta \ell_0^2$.
It is convenient
to rescale the transverse length by $\sqrt{2\pi} \ell$,
and construct a dimensionless unit cell with primitive vectors
$\br'_{\rm I}=(1/\sqrt{\eta},0)$ and
$\br'_{\rm II}=(1/2\sqrt{\eta},\sqrt{\eta})$, giving an area of unity.
A dimensionless vector $\tk$ can be defined as $\sqrt{2\pi}\ell k$.
The corresponding dimensionless BZ in the reciprocal lattice has area $4\pi^2$.
By changing $\eta$, one can construct a unit cell of different shape with
the area remaining unchanged.
The ideal triangular and square lattices correspond to $\eta_\Delta=\sqrt{3}/2$,
and $\eta_{\Box}=1/2$ respectively.
In this paper, all the integrals are evaluated at $\eta_\Delta$
except in the calculation of $c_{66}$ in Eq.\,(\ref{eqn_c66}).

In general, the function
$I(\tk_1,\tk_2|\tk_3,\tk_4)$ can be expressed in terms of gauge invariant
reciprocal lattice sums \cite{Brandt69,Yeo_private}. To simplify notation,
we will ignore the tilde assigned to the dimensionless $k$ vector
and exploit the conservation of momentum $\bk_1+\bk_2=\bk_3+\bk_4$
on each quartic vertex.

\begin{eqnarray}
I(k_1,k_2|k_3,k_4) &=&
\exp\left[-|\bk_2-\bk_4|^2/4\pi-i(k_{y2}-k_{y4})(k_{x2}-k_{x3})/2\pi
\right]   \nonumber \\
& &
\sum_{m,n}\Theta_{m,n} \exp\left[-(k_{y2}+i k_{x2})(ig_x-g_y)\right]
\nonumber \\
& & \times \exp\left[-(g_yk_{y4}+g_xk_{x4})+i(g_xk_{y3}-g_yk_{x3})\right],
\end{eqnarray}
where $(g_x,g_y)=(m\sqrt{\eta}, (n-m/2)/\sqrt{\eta})$
are dimensionless reciprocal lattice vectors,
and $\Theta_{m,n}=\exp[-\pi(g_x^2+g_y^2)]/\beta_A$.
In particular,
\begin{eqnarray}
I(0,0|0,0) &=& \sum_{m,n}\Theta_{m,n}=1\\
%
I(0,k|0,k) &=&\sum_{m,n}
\Theta_{m,n}
\exp\left[i\bigg[k_y g_x-k_x g_y \bigg]\right]\\
&\approx & 1-\left(\frac{ \eta}{4\pi\surd{3}}\right)k^2+0.0029k^4+\cdots,
~~~~~~~~~~
k \rightarrow 0 \\
|I(0,0|k,-k)| &=&\bigg| \sum_{m,n}
\Theta_{m,n}
\exp\left[\bigg(k_y-ik_x\bigg)\left(ig_x+g_y\right)-
\frac{k^2}{4\pi}\right]\bigg| \\
& \approx & 1-\left(\frac{ \eta}{2\pi\surd{3}}\right)k^2+
0.0032k^4+\cdots,~~~~~~~~~~
k \rightarrow 0.
\end{eqnarray}

Within the BZ, it can be shown that
$0<\epm<0.59<\epp<1$. The asymptotic behavior of $\epsilon_{\pm}$ can be shown to be at small $k$:
\begin{eqnarray}
\epp (k)& \approx & 2-\left(\frac{\eta}{\pi\surd{3}}\right)k^2 +0.0099k^4+\cdots\\
\epm (k)&\approx & \frac{\tilde{c}_{66}\beta_A}{(2\pi)^2} k^4
= 0.0026 k^4+\cdots.
\end{eqnarray}

\begin{figure}[htb]
\begin{center}
\mbox{\epsfxsize= 6cm
\epsfbox{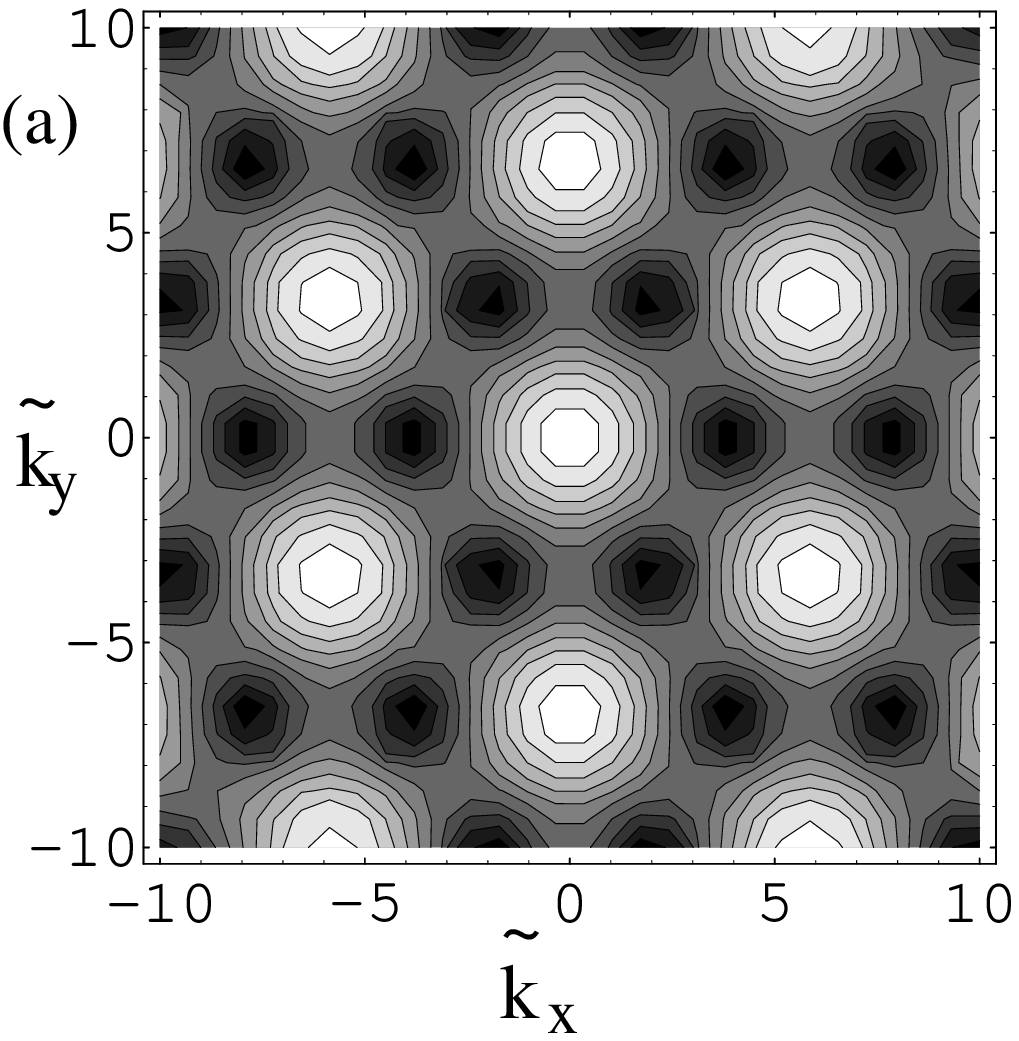}}
\mbox{\epsfxsize= 6cm
\epsfbox{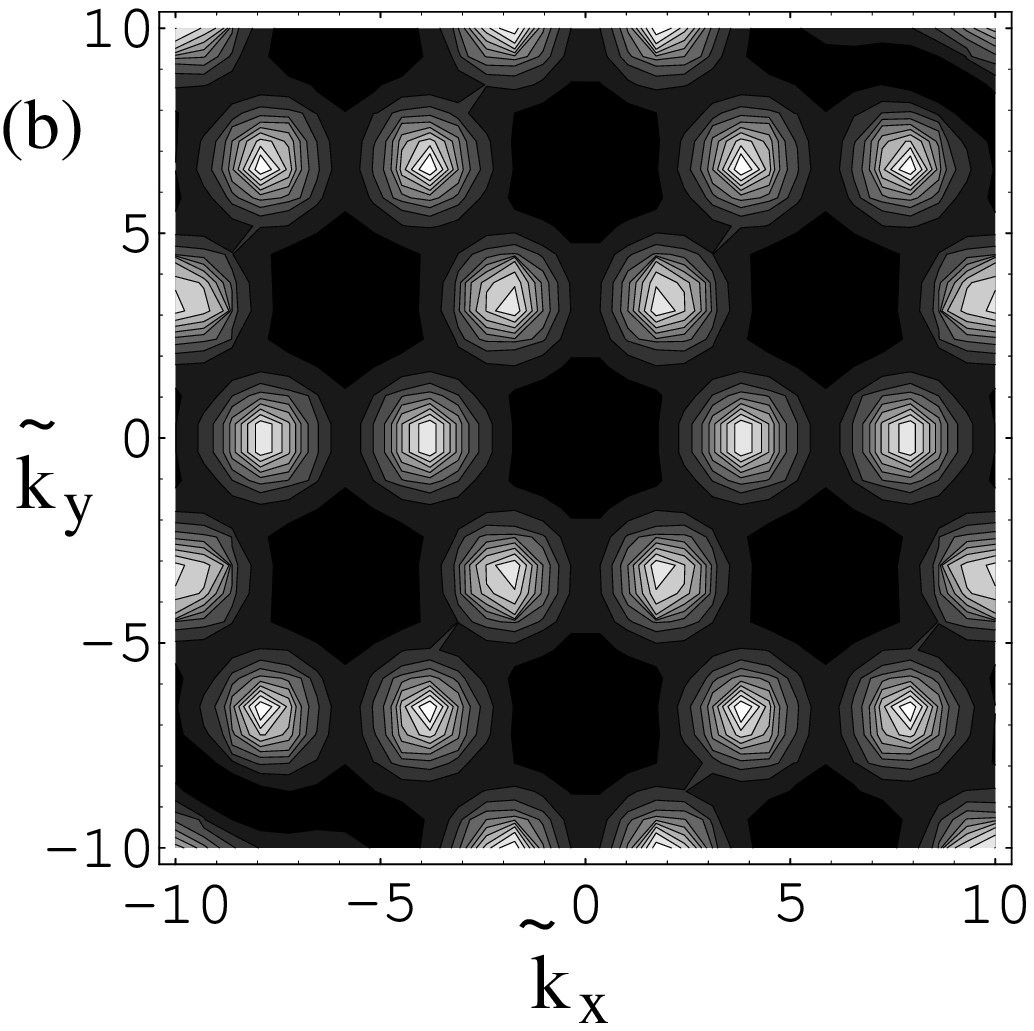}}
\caption{The contour plot of the functions (a)$\epp(\tk)$ and (b)$\epm(\tk)$
for the triangular configuration.}
\end{center}
\label{fig_tri}
\end{figure}

It is instructive to plot the contour of the function $\epsilon_{\pm}(\tk)$
for a triangular lattice
configuration to illustrate its symmetries and the first BZ
(see Fig.\,\ref{fig_tri}).
The maxima and minima are denoted by the light and dark shades respectively.
The reciprocal lattice vector(RLV) points are marked either by the maxima of $\epp$ or the
minima of $\epm$.
The equal-sided hexagonal BZ can be constructed by joining together
the six minima (maxima) surrounding the central RLV in the function $\epp$ ($\epm$).
It is easy to see that approximating the two dimensional
integration by a circular BZ underestimates the integrand
involving $\epm(\tk)$
as the function has spikes at the corners of the Brillouin zone.

\begin{thebibliography}{10}
\bibitem{decouple}
This 3D to 2D crossover is purely a finite size effect and should not be
  confused with the decoupling mechanism of the vortices mentioned earlier.
\bibitem{bc2}
The mean field coherence length $\xi^{MF}_\perp(0)$ is approximately the same
  in YBCO and BSCCO, and therefore $B_{c2}(0)\approx \Phi_0/2\pi
  \xi^{MF}_\perp(0)^2$ are also approximately the same in both cases. We have
  used a typical value of $B_{c2}(0)\approx 170T$ which coressponds to
  $\xiab^{MF}(0)\approx 14\AA$.
\end{thebibliography}

\newpage
\begincol

\endcol

\end{document}